\documentclass[10pt,twocolumn,letterpaper]{article}

\def\etal{et~al.\_}			  % and others, and co-workers
\def\eg{e.g.,~}               % for example
\def\ie{i.e.,~}               % that is, in other words
\newcommand{\bs}{\boldsymbol}

\def\sysname{IPE-LIIF\xspace}
\def\prefix{IPE\xspace}

\newlength\paramargin
\newlength\figmargin

\newlength\secmargin
\newlength\figcapmargin
\newlength\tabcapmargin

\newcommand{\lfi}[1]{\textbf{#1}}
\newcommand{\lse}[1]{{#1}}
\newcommand{\slse}[1]{#1}
\newcommand{\down}{\textcolor[RGB]{0,128,0}}

\newcommand{\topic}[1]
{
\vspace{1.5mm}\noindent\textbf{#1}
}

\long\def\ignorethis#1{}

\newbox\jsavebox%

\makeatletter
\newcommand{\providelength}[1]{%
  \@ifundefined{\expandafter\@gobble\string#1}
   {% if #1 is undefined, do \newlength
    \typeout{\string\providelength: making new length \string#1}%
    \newlength{#1}%
   }
   {% else check whether #1 is actually a length
    \sdaau@checkforlength{#1}%
   }%
}

\usepackage[pagenumbers]{cvpr} % To force page numbers, e.g. for an arXiv version

\usepackage{graphicx}
\usepackage{amsmath}
\usepackage{amssymb}
\usepackage{booktabs}
\usepackage{multirow}
\usepackage{multicol}
\DeclareMathOperator{\sinc}{sinc}

\usepackage[pagebackref,breaklinks,colorlinks]{hyperref}

\usepackage[capitalize]{cleveref}
\crefname{section}{Sec.}{Secs.}
\Crefname{section}{Section}{Sections}
\Crefname{table}{Table}{Tables}
\crefname{table}{Tab.}{Tabs.}

\def\cvprPaperID{***} % *** Enter the CVPR Paper ID here
\def\confName{CVPR}
\def\confYear{2022}

\begin{document}

\title{Enhancing Multi-Scale Implicit Learning in Image Super-Resolution with Integrated Positional Encoding}

\author{
 Ying-Tian Liu \hspace{2.5mm} Yuan-Chen Guo \hspace{2.5mm} Song-Hai Zhang \\[2mm]
 BNRist, Department of Computer Science and Technology, Tsinghua University \\
% {\tt\small liuyingt20@mails.tsinghua.edu.cn}
% \and
}
\maketitle

%%%%%%%%% ABSTRACT
\begin{abstract}
    Is the center position fully capable of representing a pixel? There is nothing wrong to represent pixels with their centers in a discrete image representation, but it makes more sense to consider each pixel as the aggregation of signals from a local area in an image super-resolution (SR) context. Despite the great capability of coordinate-based implicit representation in the field of arbitrary-scale image SR,
    this area's nature of pixels is not fully considered. To this end, we propose integrated positional encoding (IPE), extending traditional positional encoding by aggregating frequency information over the pixel area. We apply IPE to the state-of-the-art arbitrary-scale image super-resolution method: local implicit image function (LIIF), presenting \sysname. We show the effectiveness of \sysname by quantitative and qualitative evaluations, and further demonstrate the generalization ability of \prefix to larger image scales and multiple implicit-based methods. Code will be released.
\end{abstract}

%%%%%%%%% BODY TEXT
\section{Introduction}

Single image super-resolution (SISR) aims to reconstruct a visually-natural high-resolution (HR) image from its low-resolution (LR) counterpart. 
SISR is an ill-posed problem since there exist many potential reasonable HR images for a single LR input and the down-sampling filters for HR images are not determined. 
Many methods have been proposed to bridge the gap in signal frequency between LR and HR images by either regressing high-frequency textures\cite{rdn, srcnn, edsr, srdensenet, metasr} or generating perceptually reasonable details~\cite{srgan, esrgan, enhancenet, pro-perception-oriented, structure-preserving}. 
%Broadly speaking, in terms of evaluation metrics, SISR methods can be classified as PSNR-oriented~\cite{rdn, srcnn, edsr, srdensenet, metasr} and perception-oriented ones~\cite{srgan, esrgan, enhancenet, pro-perception-oriented, structure-preserving}.
Although super-resolution (SR) with predefined scale factor has been fully explored with convolution-based architectures~\cite{srcnn, srgan, lapsr, vdsr, edsr, rdn}, arbitrary-scale SISR is a more attractive and challenging task that is still under exploration. 
%few works has been done on arbitrary scale image SR, an apparently more convenient and attractive task, yet.
% MDSR~\cite{edsr} is the first attempt to simultaneously resolve the super-resolution of multiple scales with a single fusion model. 
%zsh 此处SR并没有定义，建议先定义，或者索性写全super resolution

\begin{figure}[t]
\centering
%  \resizebox{\linewidth}{!}{
\captionsetup[sub]{font=small,labelfont={bf,sf}}
\captionsetup[subfloat]{labelformat=empty,justification=centering,font=small}
 \setlength {\tabcolsep} {2pt}
\begin{tabular}{ccc}
\multicolumn{2}{c}{\multirow{-5.7}{*}{\subfloat[img0828 from DIV2K\cite{div2k} ($\times 4$)]{\includegraphics[width=0.65\linewidth,trim=342 0 342 0,clip]{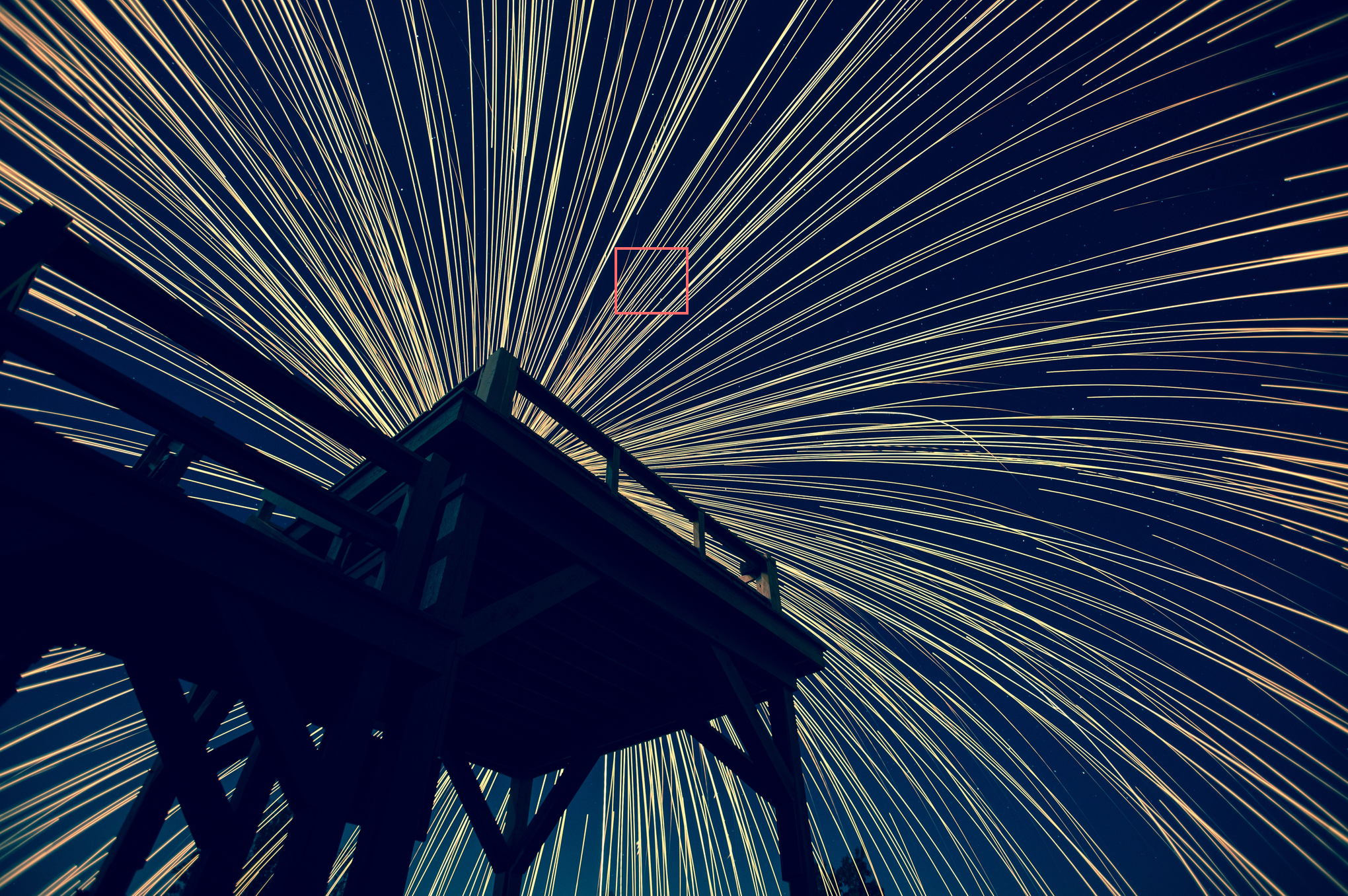}}}} & 
\subfloat[EDSR-LIIF\\(22.26dB/0.9171)]{\includegraphics[width=0.3\linewidth]{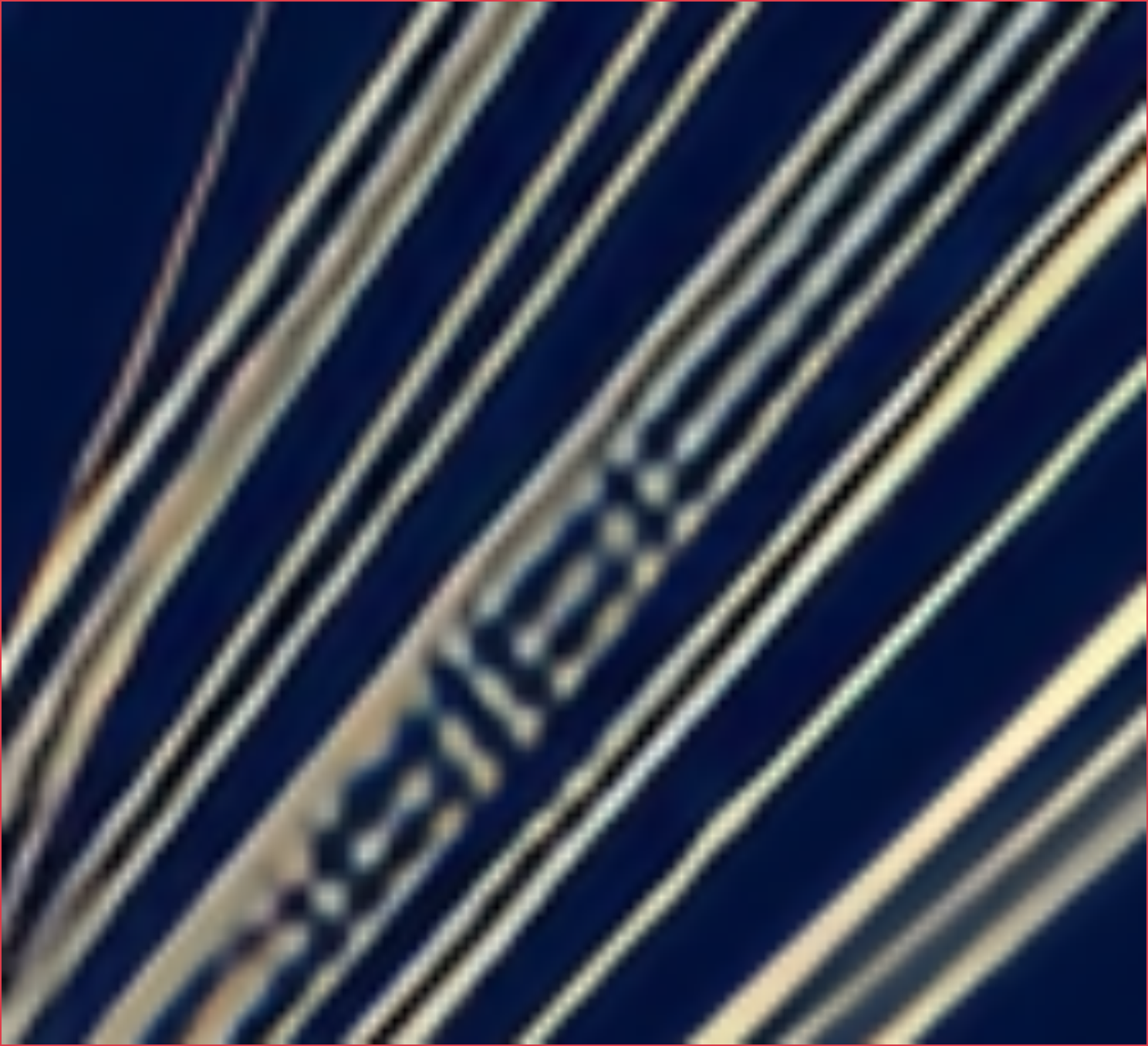}} \\
\multicolumn{2}{c}{} & 
\subfloat[\textbf{EDSR-IPE-LIIF\\(22.41dB/0.9242)}]{\includegraphics[width=0.3\linewidth]{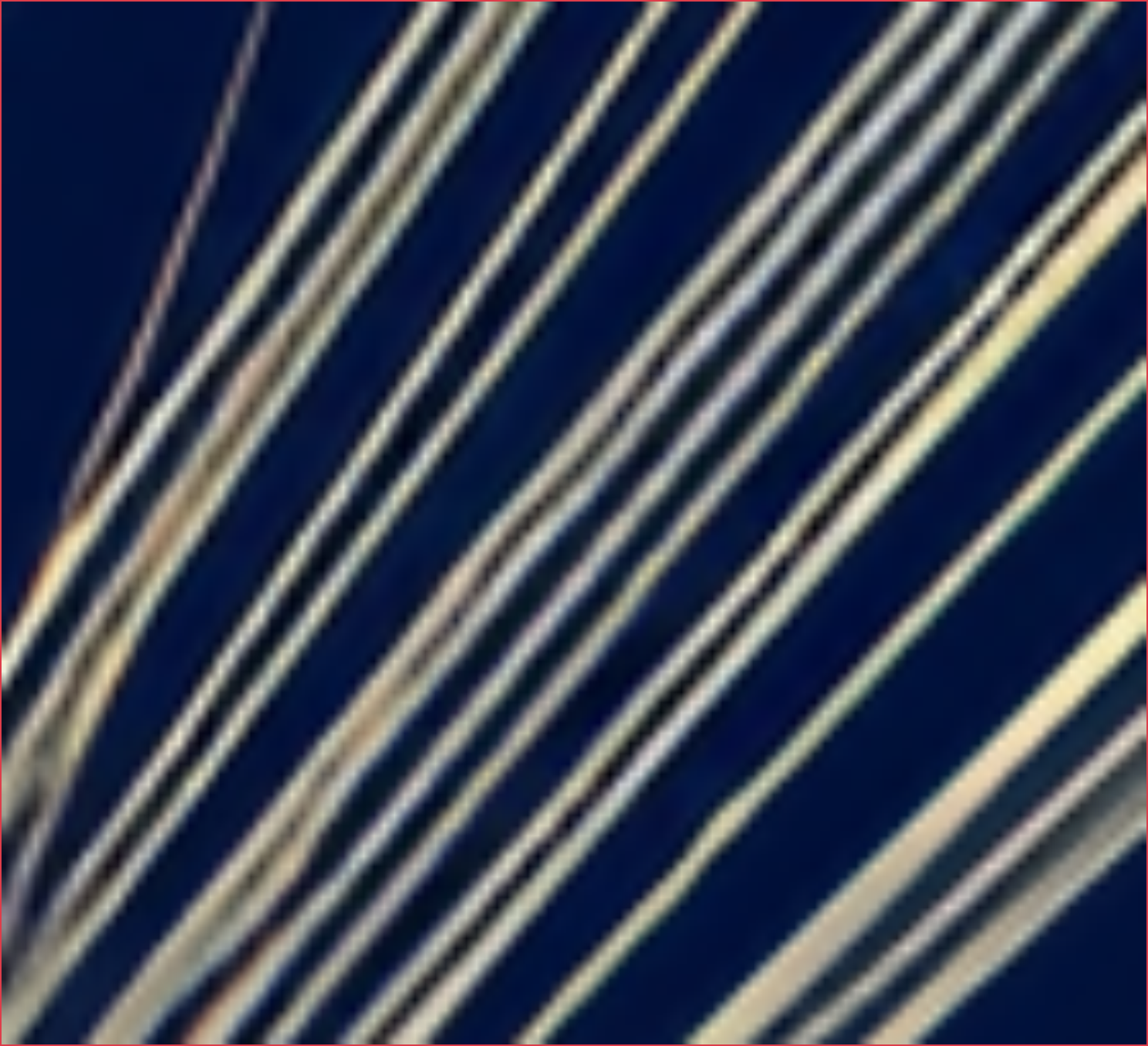}} \\
% 22.26 22.41 23.42 23.83 

% & \subfloat[EDSR-IPE-LIIF\\(29.30dB)]{\includegraphics[width=0.15\linewidth]{figure/result/cropped/set14-2-edsr-ipe_0001_pred_magnifier_0.png}} \\
% &  \subfloat[Bicubic\\(27.94dB)]{\includegraphics[width=0.15\linewidth]{figure/result/cropped/X2_0001_pred_magnifier_0.png}} 
 \subfloat[RDN-LIIF\\(23.42dB/0.9357)]{\includegraphics[width=0.3\linewidth]{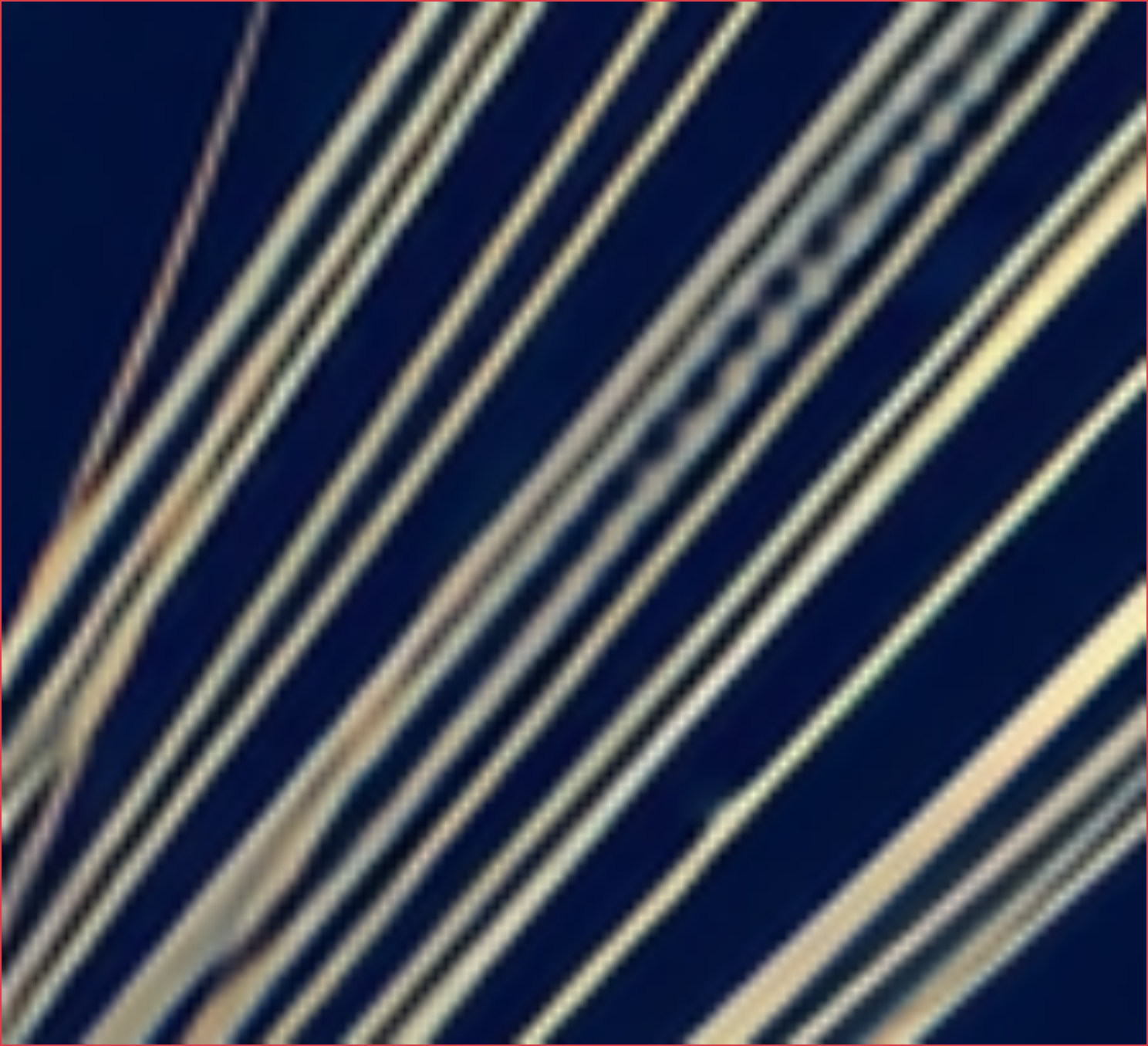}} 
&  \subfloat[\textbf{RDN-IPE-LIIF\\(23.83dB/0.9420)}]{\includegraphics[width=0.3\linewidth]{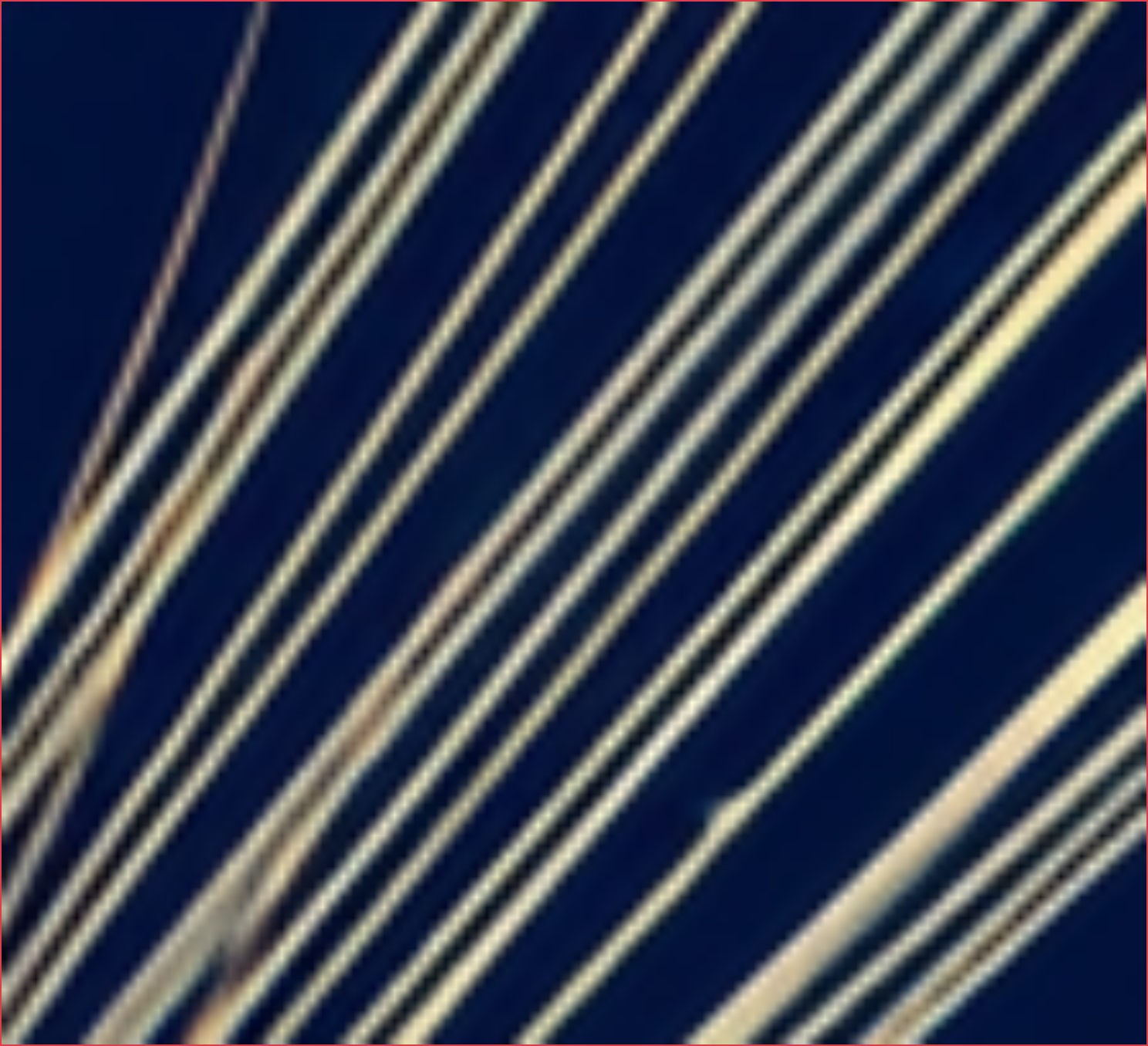}}
&
\subfloat[HR\\(PSNR/SSIM)]{\includegraphics[width=0.3\linewidth]{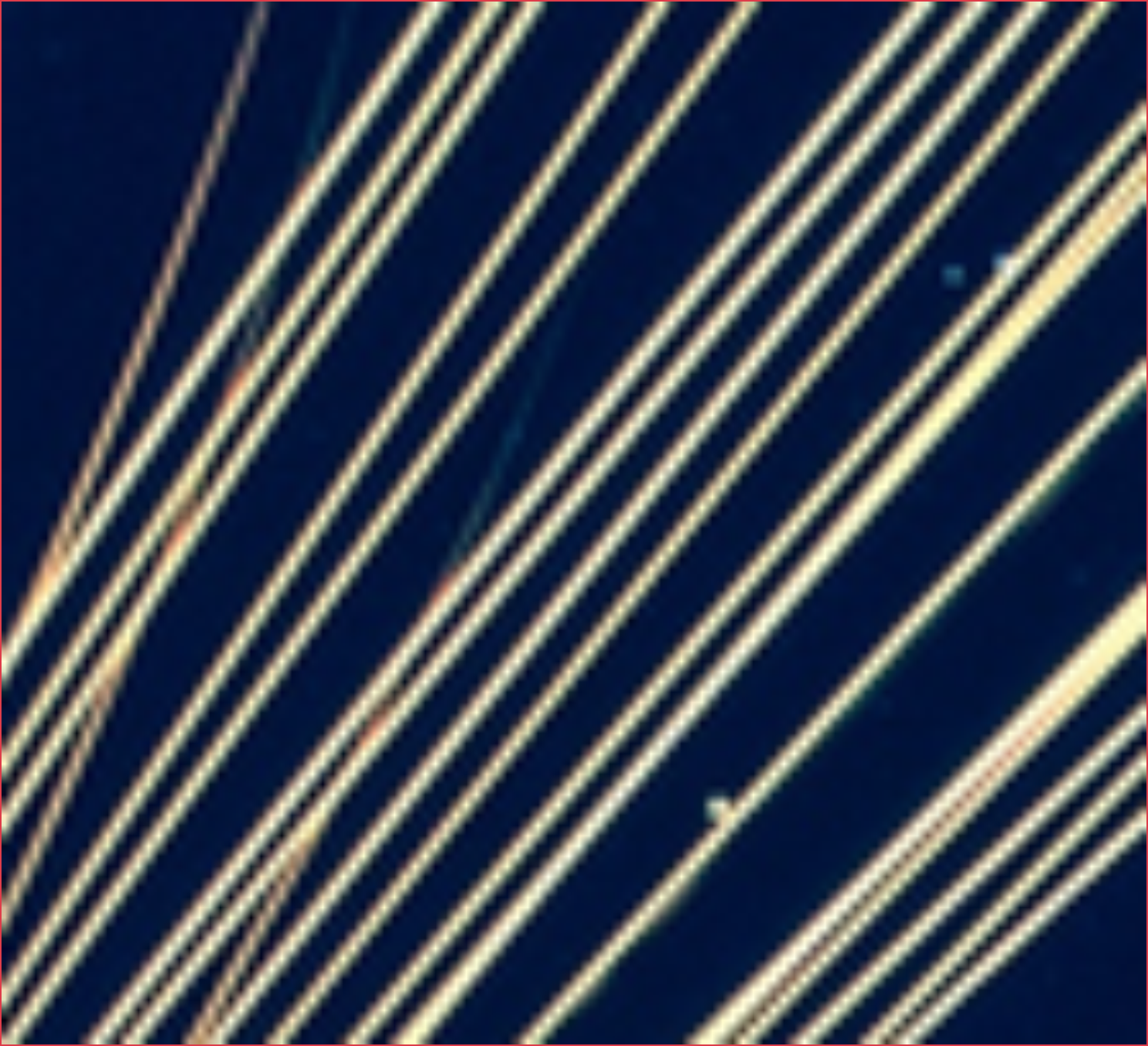}} \\

\end{tabular}
% }
\caption{$\times 4$ SR results of methods using the integrated positional encoding (IPE) compared with those without IPE.}
\label{fig:teaser_example}
\vspace{-4mm}
\end{figure}

Recently, implicit neural representations have emerged to represent various kinds of signals. The key idea is to represent a signal as a function parameterized with a multi-layer perceptron (MLP) that maps a coordinate input to the signal value at that position. What makes the neural implicit representation appealing is that a signal is often represented continuously or in infinite resolution and contains sufficient detail. 
%An implicitly represented signal is often continuous or in infinite resolution since the input coordinates can be arbitrarily fine, making the neural implicit representation quite appealing. 
Different local signals could form various kinds of entities on the whole, such as density and radiance field for 3D scenes~\cite{nerf, mip-nerf}, signed distance field for 3D shapes~\cite{acorn,shape-templates, bspnet, deepsdf, imnet} and RGB values for images/videos~\cite{acorn, siren, fourier-features}.

Based on implicit neural representations, Meta-SR\cite{metasr} achieved arbitrary-scale image SR by generating variant weights of the up-scaling filters for a position and scale input.
Chen~\etal~\cite{liif} proposed the local implicit image function (LIIF), for continuously representing natural and complex images.
LIIF learns an implicit decoder across images to predict the color within the continuous neighborhood of a feature vector in a 2D feature map. 
% Each ``continuous" image is represented as a 2D feature map.
%whose latent vectors are evenly distributed within the image range. 
% Each feature vector located at discrete positions of the feature map is responsible for color prediction in its continuous neighborhood with a learnable implicit decoder shared across images. The decode takes the relative position of the pixel center and the local feature vector as input and outputs the color value. 
Therefore, LIIF builds a bridge between the discrete and continuous representation for 2D images. Both Meta-SR\cite{metasr} and LIIF\cite{liif} exploit the convenience of implicit representation.

% However, the continuity of the MLP makes it difficult for small changes in the input to be reflected in the output, which may introduce bias in the spectrum\cite{spectralbias}. 

However,  MLPs in implicit representation have difficulty in learning high-frequency functions, which is also known as the spectrum bias~\cite{spectralbias}. The bias hinders the generation of 
%the difficulty in fitting high-frequency signals or producing 
detail-rich contents in some scenarios, such as image SR and novel view synthesis. 
% Difficult fitting of high-frequency signals makes MLP hardly meet the requirements of generating detail-rich contents in some scenarios, such as image super-resolution and novel view synthesis. 
NeRF~\cite{nerf}, a novel framework for novel view synthesis, adopts positional encoding (PE) to address this issue. PE allows MLPs to spatially distinguish inputs with small differences to recover high-frequency details. But in NeRF's rendering, each pixel emits only one ray to predict its color, which may lead to observable aliasing artifacts. To conform as much as possible to the physical aggregation of light, an approximate integral of PE along the ray cone is used in \cite{mip-nerf} to make full use of multi-scale training images and improve the rendering efficiency of anti-aliased views.

In SISR, the same problem exists. Imagine that we have an HR image and a corresponding LR image of half the size, where the pixels in the LR image are not necessarily the same as the pixels at even positions in the HR image. To address the problem, LIIF\cite{liif} directly inputs the pixel size to the implicit decoder to distinguish them, called \textbf{cell decoding}. However, the direct input of coordinates and size makes it difficult for the implicit decoder to fit high-frequency details.
% Without PE, LIIF~\cite{liif} relies solely on the pixel location and size, making it easier to produce images with missing details as the magnification increases.
Although UltraSR~\cite{ultrasr} has noticed the problem and integrates PE to improve the SR performance, it still does not consider multi-scale variations simultaneously, which is critical in tasks with the output of different pixel sizes. We believe that each pixel in the SR task should be treated as the aggregation of colors over a small area, instead of being represented exactly by its center location. Both pixel size and position are critical for high-quality arbitrary-scale SR. For this purpose, we propose IPE, which is to compute the integral of PE over the pixel to be predicted. We then present \sysname, where the IPE instead of the coordinate input is concatenated with the local feature vector as the input of the implicit decoder. Experiments reveal that \sysname achieves state-of-the-art performance on several popular SR datasets, and multiple methods adopting IPE also give consistently better results than the original version when the SR scales varies over a wider range.
% Mip-NeRF\cite{mip-nerf} holds a similar view to ours in terms of improvements to NeRF. In mip-NeRF, an approximate integral of PE encoding the shape and spatial location of ray cone is proposed to replace the plain PE, significantly improving the scene quality with multi-scale input images. 
The contributions of this paper are threefold:
\begin{itemize}
    \item We propose integrated positional encoding (IPE), which can provide scale and position-dependent encoding for implicitly represented pixels in image SR. 
    \item The arbitrary scale SR model incorporating IPE has demonstrated its optimal competence in both qualitative and quantitative experiments.
    \item Further experiments highlight that IPE can be applied to a variety of SR models based on implicit representations, consistently improving their capacity to learn multi-scale outputs.
\end{itemize}

% The solution we present, which we call integrated positional encoding (\prefix), improves the implicit learning capability of existing methods for multi-scale samples. IPE will replace the coordinate input with the expected positional encoding over the pixel region where the color signals should be integrated. IPE is shown to be able to improve the performance of multiple implicit-based SR methods, such as Meta-SR\cite{metasr} and LIIF\cite{liif} in learning multi-scale outputs.
% Experiments highlight that \sysname combining \prefix and local implicit image function achieves the state-of-the-art performance for the arbitrary scale SR task.
\section{Related Work}

% Our work has been improved based on LIIF\cite{liif}, achieving the state-of-the-art performance on the arbitrary scale image super-resolution task. LIIF\cite{liif} introduced an arbitrary scale image super-resolution technique that could derive the pixel value at any position from a rasterized feature map with an implicit decoder. 

In this section, we will review the SISR task and the implicit representation with a focus on positional encoding (PE).

\topic{Single image super-resolution}
SISR is one of the most popular tasks in the visual community\cite{anchored, neighbor-embedding, srcnn, espcnn, rcan, dbpn, vdsr, memnet, srdensenet, deep-unfolding}. Traditional methods\cite{anchored, neighbor-embedding} are mainly example-based. 
SRCNN\cite{srcnn} is the first end-to-end CNN-based SR model. It implements the process of feature extraction, nonlinear mapping, and image reconstruction with a linear three-layer structure. Recursive strategies\cite{drcn, drrn} were proposed to improve the training stability and detail fidelity. VDSR\cite{vdsr} and IRCNN\cite{ircnn} made the network deeper by stacking convolutional layers with residual learning.  For the first time, ESPCNN\cite{espcnn} proposed to compute on low-resolution inputs before the up-sampling layer, significantly reducing computational overhead.
SRResNet\cite{srgan} introduced the residual connection\cite{resnet} into image SR and generated perceptually sound details via adversarial training. EDSR\cite{edsr} improved the ResNet block by removing the BatchNorm\cite{bn} layer. Moreover, it proposed to learn simultaneously the backbone corresponding to various SR scales, forming the fusion model MDSR\cite{edsr}, which allowed multi-scale SR. Similarly, 
LapSR\cite{lapsr} proposed a progressive up-scaling architecture and a novel hierarchical reconstruction loss to complete SR with scales for powers of 2. With the observation that the dense connection\cite{densenet} allows feature reuse from all the preceding layers within each block, RDN\cite{rdn} adopted stacked dense blocks to make a full use of hierarchical features extracted from LR image, thereby achieving relatively-high performance. 

\topic{Arbitrary scale super-resolution}
Image SR on an arbitrary scale factor is obviously more attractive and convenient than that on a pre-defined one. Hu \etal\cite{metasr} proposed to learn a meta-upscale module to dynamically predict the weights of the up-scaling filters on different scales. Meta-SR\cite{metasr} model could combine the meta-upscale module with the feature learning module from most previous methods\cite{rdn, edsr, rcan}, producing convincing SR results on arbitrary scale. 
LIIF \cite{liif} replaced the meta-upscale module with an implicit decoder, representing natural and complex images in a continuous manner. And it is proved to be effective for image SR at out-of-distribution scales, yielding visually smoother results than Meta-SR\cite{metasr}.

\topic{Coordinate-based neural representation}
Neural implicit representations can faithfully reconstruct various signals and have therefore received extensive attention and research\cite{occupancynet, deepsdf, imnet, nerf}. An object in implicit representation can be typically modeled as a multi-layer perceptron (MLP), mapping a coordinate input to the signal at that 2D/3D location. Such an idea has shown effectiveness in a variety of scenarios, \eg 3D shape modeling\cite{imnet, shape-templates, bspnet, sal}, neural volume rendering\cite{nerf, mip-nerf, nsvf}, 3D scene modeling\cite{implicitgrid, scene3d} and image/video representation\cite{siren, liif}. A recent work\cite{spectralbias}  has highlighted that there exists a spectral bias in the MLP activated by ReLU, which leads to lack of details in  high-frequency signal representation. This problem can be solved by either replacing ReLU with periodic functions\cite{siren} or adding an extra PE of the coordinate input\cite{nerf, mip-nerf, nsvf, fourier-features, ultrasr}. The use of proper PE can significantly 
accelerate the training process and improve the fidelity of signals at high-frequency positions. UltraSR\cite{ultrasr} adds spatial encoding into LIIF\cite{liif}, reducing checkboard artifacts on the large scale SR. Inspired by the pre-computed multi-scale anti-aliasing technique (\eg mipmap\cite{mipmap}), Mip-NeRF\cite{mip-nerf} accelerated the anti-alias novel view rendering in NeRF with an approximate integral of PE within the interest region. 

% Pixels can not be represented by coordinates exactly. 
The method we propose, named \prefix, provides position and shape information in a uniform encoding for each pixel. \sysname applys \prefix to LIIF, achieving the state-of-the-art performance in the arbitrary-scale SR task. We verify the effectiveness and generality of our design with extensive experiments and convincing ablation studies. 

s
\begin{figure*}[htbp]
    \centering
    \includegraphics[width=0.9\linewidth]{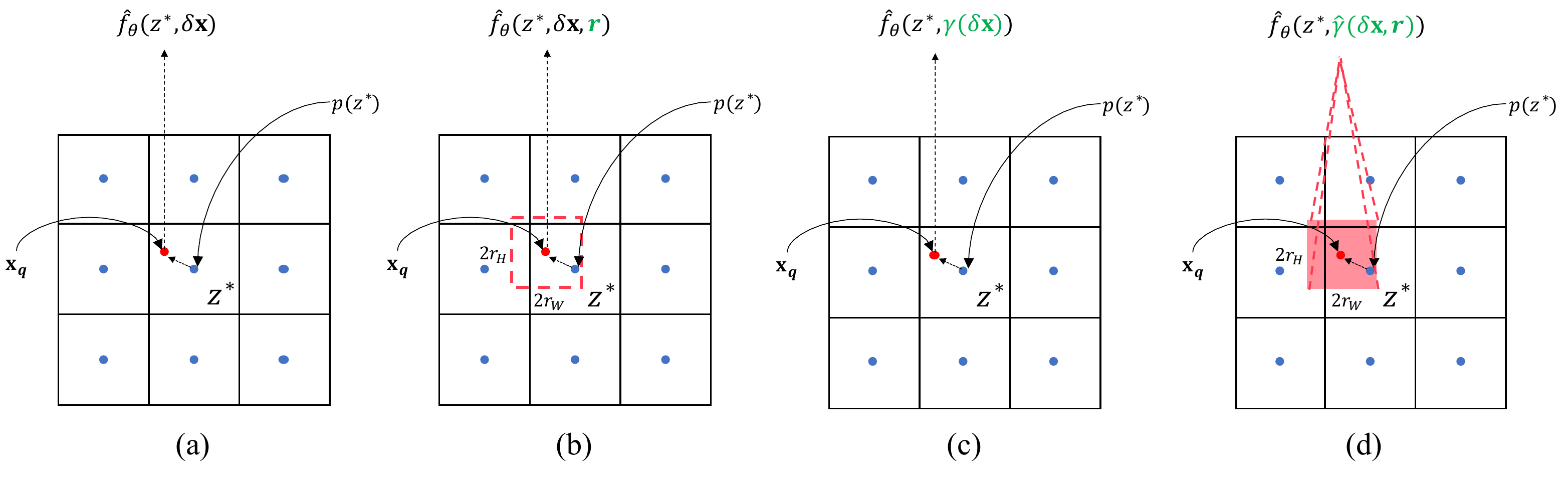}
    \caption{Illustrations of different spatial encodings on this $3 \times 3$ feature map: (a) no encoding, (b), cell decoding used in \cite{liif}, (c) plain PE used in\cite{nerf, ultrasr}, (d) IPE (ours).  $\mathbf{x}_q$ denotes the center of the query pixel. The query pixel is marked with a red box in (b) and (d), with the radius $(r_W, r_H)$. The differences in the expressions are marked in green. }
    \label{fig:encoding}
    \vspace{-4mm}
\end{figure*}

\section{Preliminaries: local implicit image function}
\label{sec:preliminaries}
LIIF\cite{liif} performs the arbitrary-scale image SR by learning a continuous representation for images. In continuous representation, each image $I$ is represented as a 2D feature map $M \in \mathbb{R}^{H \times W \times C}$, where the size is $H \times W$ and the channel is of length $C$. To predict the RGB value at an arbitrary position, LIIF uses a learnable implicit function, which takes the coordinate and the nearest feature as the input and predicts the RGB value. The implicit function is parameterized by an MLP $f_{\theta}$(with $\theta$ as its parameter) and has the following form:
\begin{equation}
    s = f_{\theta}(z, \mathbf{x})
    \label{eq:implicit-decoder}
\end{equation}
where $\mathbf{x}$ is a continuous 2D coordinate in the range $\chi = [0,2W]\times[0,2H]$ and $z$ is a latent vector representing the local feature. The output $s$ is the predicted color at position $\mathbf{x}$. With a pre-defined $f_{\theta}$, $f_{\theta}(z, \cdot)$ becomes a local implicit function representing the local image area.

In practice, the $H \times W$ features are evenly distributed in $\chi$. The RGB value of the continuous image $I$ at a query coordinate $\mathbf{x}_q$ is 
\begin{equation}
    I(\mathbf{x}_q) = \hat{f}_{\theta}(z^*, \mathbf{x}_q - p(z^*))
    \label{eq:practice}
\end{equation}
where $z^*$ is the nearest feature vector to position $\mathbf{x}_q$ in $M$, $p(z)$ is the position of $z$. Note that the implicit function $\hat{f}_{\theta}$ here describes the location relatively, which is different from $f_{\theta}$ defined in \cref{eq:implicit-decoder} with absolute coordinate input . 
In \cref{eq:practice}, the relative coordinate $\delta\mathbf{x} = \mathbf{x}_q -p(z^*)$ is in the range of $[-1, 1]^2$. Each latent code $z$ in $M$ is responsible for predicting the color in such a local area around $p(z)$. With the shared implicit decoder, LIIF bridges the gap between features at discrete positions and a continuous image. 

With this procedure for rendering the continuous image at the arbitrary resolution, training LIIF is straightforward: sampling pixels from HR images to build coordinate-color pairs, we minimize the difference between the predicted RGB values and the ground truths with gradient descent. Since LIIF works as a decoder that converts the feature map into an image, it is usually jointly trained with a feature extraction module in CNN models\cite{edsr, rdn}.

\section{Method}
In this section, we first discuss the shortcomings of LIIF\cite{liif}, propose the solution, and finally present the overall architecture of our model. As described in \cref{sec:preliminaries}, LIIF\cite{liif} samples only the center to render each pixel. Essentially, LIIF ignores the fact that an image pixel is actually the aggregation of the color in a small area, and considers the color value at the central point as the color of that pixel. Imagine that we have an HR image and a corresponding LR image down-sampled by 2, where pixels in the LR image are not necessarily the same as those at even positions in HR. Therefore the scaling factor or, in other words, the size of the target pixel is crucial in this task. And to address the problem, the original LIIF\cite{liif} simply concatenates the pixel size and the central position of pixel area as input to the MLP, called \textbf{cell decoding}. But both our and LIIF's experiments reveal that this cell decoding strategy is not guaranteed to improve the quality of results. 

\begin{figure}[htbp]
    \centering
    \includegraphics[width=\linewidth,clip, trim={0 10cm 0 10cm}]{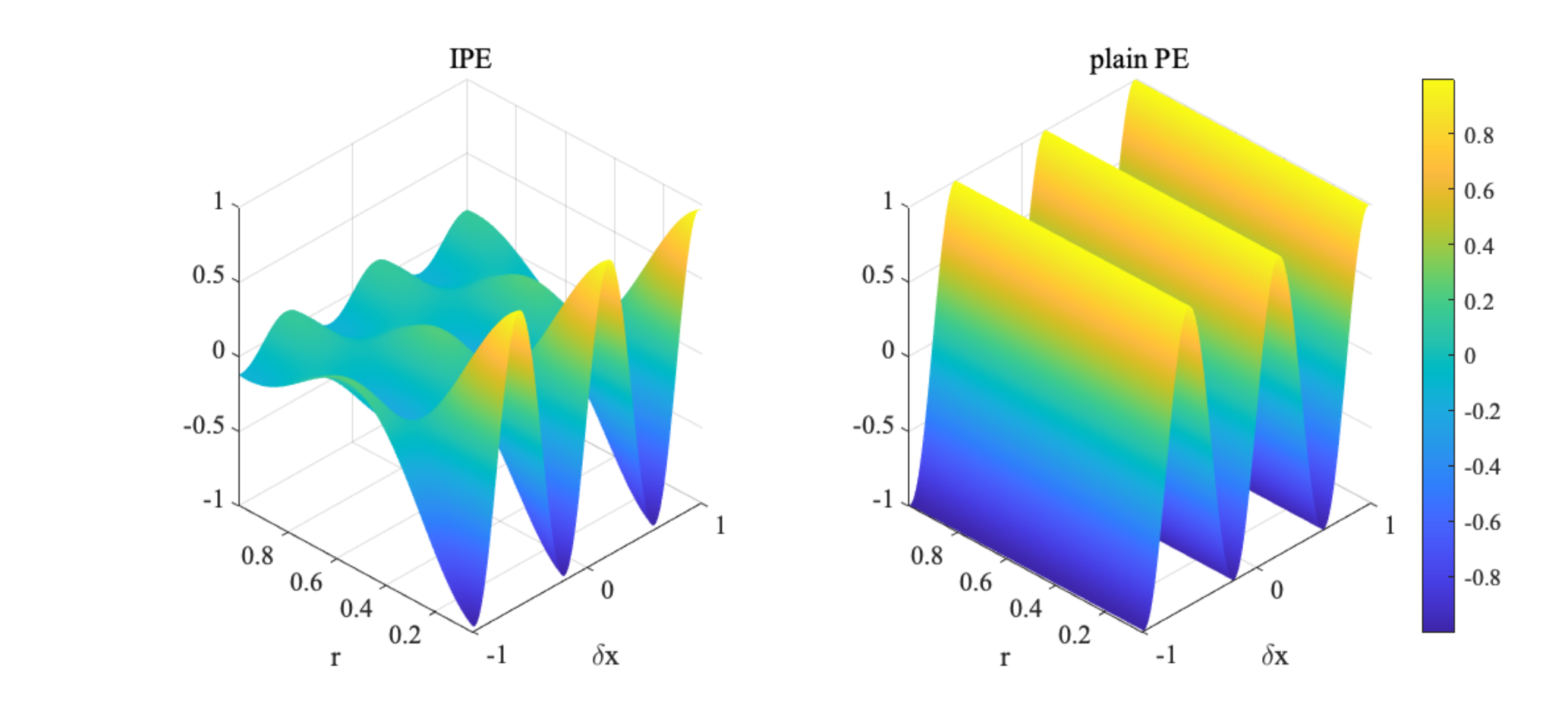}
    \caption{Comparison between the sine encoding of IPE (ours) and that of plain PE with the frequency $\omega = 8$.}
    \label{fig:pe}
    \vspace{-4mm}
\end{figure}

\subsection{Integrated Positional Encoding}

We solve the problem by introducing IPE of each pixel, like \cite{mip-nerf}, enabling the implicit decoder to focus on the local region rather than a center point without area. Moreover, with IPE, the implicit decoder could adapt to a variety of SR scales, which are essentially related to the pixel size to decode.

The commonly used PE\cite{nerf, fourier-features} transforms each position with periodic function in different frequencies:
\begin{equation}
    \gamma(\mathbf{x}) = [\sin(\mathbf{x}), \cos(\mathbf{x}), ..., \sin(2^{L - 1}\mathbf{x}), \cos(2^{L-1}\mathbf{x})]
\end{equation}

Similar with \cite{mip-nerf}, we construct the featurized representation of the pixel with the expected PE of all the coordinates that lie in it. The main differences among several spatial encoding methods are illustrated in \cref{fig:encoding}. In \cite{mip-nerf}, the integral of PE in a conical frustum is quite tricky and has no closed form solution, so they use a multivariate Gaussian with the same mean and variance to approximate the area. In the 2D case we are discussing here, the integral is straightforward:
within a pixel $l$ whose center is $\bs{c}(c_x, c_y)$ and radius (half of the edge) is $\bs{r}(r_W, r_H)$, the sine encoding with frequency $\omega$ has an expectation as:
\begin{equation}
\begin{aligned}
    \mathbb{E}_{\mathbf{x} \in l}[\sin(\omega\mathbf{x})] & = \frac {1} {4r_Wr_H}\iint_{l} \sin(\omega\mathbf{x})dxdy \\
    & = \frac {\sin(\omega\bs{c})\sin(\omega\bs{r})} {\omega\bs{r}} \\
    & = \sin(\omega\bs{c})\sinc(\omega\bs{r})
\end{aligned}
\end{equation}
Similarly, the cosine encoding is as follows:
\begin{equation}
\begin{aligned}
    \mathbb{E}_{\mathbf{x} \in l}[\cos(\omega\mathbf{x})] & = \frac {1} {4r_Wr_H}\iint_{l} \cos(\omega\mathbf{x})dxdy \\
    % & = \frac {\cos(\omega\bs{c})\sin(\omega\bs{r})} {\omega\bs{r}} \\
    & = \cos(\omega\bs{c})\sinc(\omega\bs{r})
\end{aligned}
\end{equation}
Hence, the pixel $l$ defined by its center $\bs{c}$ and radius $\bs{r}$ has the following IPE:
\begin{equation}
\begin{aligned}
\hat{\gamma}(\bs{c}, \bs{r}) = \frac 1 {\bs{r}}[\sin(\bs{c})\sin(\bs{r}), \cos(\bs{c})\sin(\bs{r}), ..., \\ 
\frac {\sin(2^{L-1}\bs{c})\sin(2^{L-1}\bs{r})} {2^{L-1}}, \frac {\cos(2^{L-1}\bs{c})\sin(2^{L-1}\bs{r})} {2^{L-1}}]
\label{eq:ipe}
\end{aligned}
\end{equation}

Intuitively, the IPE contains both location and size information, which are modulated by sine/cosine and sinc function, respectively. When the SR scale is small, the sinc part makes plain PE smoother, in which case the predicted pixel can inherit enough information from the feature vector. As the SR scale increases, the sinc part will move towards 1, making the IPE closer to plain PE (see \cref{fig:pe} for brief comparison). In that case, each feature vector is responsible for a large number of pixels, which is roughly the square of the scale. Therefore, the implicit decoder cannot reconstruct a detail-rich image solely from the features. The IPE will provide high-frequency encoding of the location to compensate for the lack of information here. The choice of hyper-parameter $L$ will further be discussed in \cref{sec:ablation}.

\begin{figure}[htbp]
    \centering
    \includegraphics[width=0.6\linewidth]{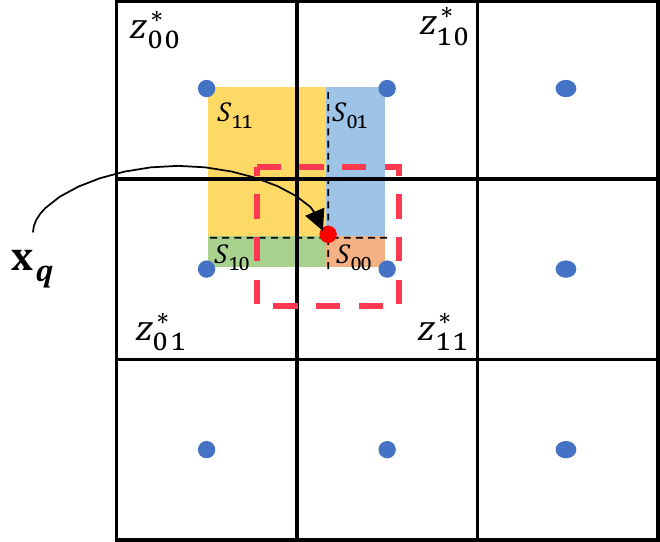}
    \caption{Local ensemble in \sysname. The query pixel along with its center is marked in red. The RGB value of the query pixel is obtained by bilinear interpolation of the results predicted by its surrounding feature vectors.}
    \label{fig:local_ensemble}
    \vspace{-4mm}
\end{figure}

\begin{table*}[tb]
    \centering
    \resizebox{\linewidth}{!}{
     \setlength {\tabcolsep} {2pt}
    \begin{tabular}{r|ccc|ccccc}
    \toprule
    \multirow{2}{*}{Method} & \multicolumn{3}{c|}{In-distribution} & \multicolumn{5}{c}{Out-of-distribution} \\
       & $\times 2$ & $\times 3$& $\times 4$ & $\times 6$& $\times 12$& $\times 18$& $\times 24$ & $\times 30$\\
    \midrule
    Bicubic\cite{edsr}& 31.01/0.9393 & 28.22/0.8906 & 26.66/0.8521 & 24.82/0.8014 & 22.27/0.7354 & 20.50/0.6986 & 19.79/0.6817 & 19.43/0.6711 \\
    EDSR-baseline\cite{edsr}& 34.55/0.9671 & 30.90/0.9298 & 28.94/0.8962 & - & - & - & - & -\\
    EDSR-baseline-MetaSR\cite{metasr} & 34.64/0.9674 & 30.93/0.9301 & 28.92/0.8962 & 26.61/0.8442 & 23.55/0.7652 & 22.03/0.7285 & 21.06/0.7058 & 20.37/0.6903 \\
    EDSR-baseline-LIIF\cite{liif} & \lse{34.67}/0.9675 & \lse{30.96}/0.9306 & \lse{29.00}/0.8974 & \lse{26.75}/0.8477 & \lse{23.71}/0.7724 & \lse{22.17}/0.7369 & \lse{21.18}/0.7137 & \lse{20.48}/0.6977\\
    % EDSR-baseline-\sysname(ours) & \lfi{34.71} &	\lfi{31.00} & \lfi{29.03} &	\lfi{26.78}	& \lfi{23.71} & 	\lfi{22.19} & \lse{21.17} & \lfi{20.48}\\
    EDSR-baseline-\sysname(ours) & \lfi{34.72}/\lfi{0.9678} & 	\lfi{31.01}/\lfi{0.9310}	& \lfi{29.04}/\lfi{0.8979}	& \lfi{26.79}/\lfi{0.8483} & \lfi{23.75}/\lfi{0.7731} & \lfi{22.21}/\lfi{0.7374} & \lfi{21.22}/\lfi{0.7143}	& \lfi{20.51}/\lfi{0.6979} \\
    \midrule
    
    RDN-MetaSR\cite{metasr} & \lse{35.00}/0.9692 & \lse{31.27}/0.9340 & 29.25/0.9016 & 26.88/0.8508 & 23.73/0.7721 & 22.18/0.7337 & 21.17/0.7104 & 20.47/0.6943 \\
    RDN-LIIF\cite{liif} & 34.99/0.9691 & 31.26/0.9339 & \lse{29.27}/0.9017 & \lse{26.99}/0.8528 & \lse{23.89}/0.7771 & \lse{22.34}/0.7412 & \lse{21.31}/\lfi{0.7181} & \lse{20.59}/\lfi{0.7013} \\
    RDN-\sysname(ours) & \lfi{35.04}/\lfi{0.9694} & \lfi{31.32}/\lfi{0.9343}	&\lfi{29.32}/\lfi{0.9025}	& \lfi{27.04}/\lfi{0.8537} & \lfi{23.93}/\lfi{0.7778} & \lfi{22.38}/\lfi{0.7415} & \lfi{21.34}/0.7177	& \lfi{20.63}/0.7011 \\
    
    \bottomrule
    \end{tabular}
    }
    \caption{Quantitative comparison on DIV2K\cite{div2k} validation set with PSNR (dB) / SSIM on different scales. Bicubic, MetaSR\cite{metasr}, LIIF\cite{liif} uses a uniform model/algorithm for all the scales. EDSR\cite{edsr} uses different models for different scales and cannot be evaluated on out-of-distribution scales. The best results under different settings are bolded. }
    \label{tab:quantitative-div2k}
    \vspace{-4mm}
\end{table*}

\subsection{Architecture details}
Given a pixel-based image $I \in \mathbb{R}^{H\times W \times 3 }$, we extract the 2D feature map $M \in \mathbb{R}^{H \times W \times C}$ with a feature extraction module $\mathcal{E}$. The module can inherit from EDSR-baseline\cite{edsr} and RDN \cite{rdn}. Feature unfolding in \cite{liif} is preserved. Each latent vector in the new feature map $\hat{M} \in \mathbb{R}^{H \times W \times 9C}$ is the concatenation of the $3 \times 3$ local neighboring vectors from zero-padded $M$.
The feature extraction process is formulated as:
\begin{equation}
\begin{aligned}
    M & = \mathcal{E}(I) \\
    \hat{M}_{ij} & = Concat(\{M_{i + k, j + m}, k, m \in \{-1, 0, 1\}\})
\end{aligned}
\end{equation}
To make the predicted RGB value smoother, we apply the local ensemble as \cite{liif}. That is, given a SR scale $s$ and a query coordinate $\mathbf{x}_q$, we extend \cref{eq:practice} to 
\begin{equation}
    I_{s}(\mathbf{x}_q) = \sum_{t\in \{00, 01, 10, 11\}}\frac {S_t} S \cdot \hat{f}_{\theta}(z_t^*, \hat{\gamma}(\mathbf{x}_q - p(z_t^*), (\frac 1 s, \frac 1 s)))
\end{equation}
where $\{z_t^*\}$ are four feature vectors surrounding $\mathbf{x}_q$ in $\hat{M}$, and they are on the vertices of a square. $S$ is the area of this square and  $S_t$ is the area of the rectangle inside this square that is opposite to $p(z_t^*)$ (\cref{fig:local_ensemble}). The query pixel radius is $(\frac 1 s, \frac 1 s)$ here if we specify the range of the $xy$-axis of the per-pixel local coordinate system in the original input LR image to $[-1, 1]$. 

The implicit decoder $\hat{f}_{\theta}$ is implemented as an MLP with skip connections\cite{nerf, deepsdf} from the input to all hidden layers. The MLP has 4 hidden layers, each with a width of 256. The bandwidth parameter $L$ of IPE is set to 10. To fuse the spatial encoding $\hat{\gamma}(\bs{c}, \bs{r})$ and the feature vector $z^*$, we concatenate them as the input of $\hat{f}_{\theta}$ (\cref{fig:encoding} (d)).

\begin{table*}[tb]
    \centering
    % \footnotesize
    \resizebox{0.8\linewidth}{!}{
    \begin{tabular}{c|c|ccc|cc}
    \toprule
    \multirow{2}{*}{Dataset} & \multirow{2}{*}{Method} & \multicolumn{3}{c|}{In-distribution} & \multicolumn{2}{c}{Out-of-distribution} \\
     & &  $\times 2$ & $\times 3$& $\times 4$ & $\times 6$& $\times 8$\\
    \midrule
    \multirow{4}{*}{Set5\cite{set5}} & RDN\cite{rdn} & \lfi{38.24}/\lfi{0.9614} & \lfi{34.71}/\lfi{0.9296} & 32.47/\lfi{0.8990}  & - & - \\
                          & RDN-MetaSR\cite{metasr} & \lse{38.10}/0.9603 & 34.62/0.9279 & 32.37/0.8959 & 28.98/0.8262 & 26.95/0.7641 \\
                          & RDN-LIIF\cite{liif} & 38.07/0.9601 & 34.63/0.9280 & \lse{32.47}/0.8969 & \lse{29.13}/0.8327 & \lse{27.12}/0.7766 \\
                        %   & RDN-\sysname (ours) & 38.20 & \lfi{34.71} & \lfi{32.53} &  \lfi{29.25} & \lfi{27.20} \\
                          & RDN-\sysname (ours) & 38.11/0.9604 & 34.68/0.9285 & \lfi{32.51}/0.8975 &  \lfi{29.25}/\lfi{0.8344} & \lfi{27.22}/\lfi{0.7797} \\
    \midrule
    \multirow{4}{*}{Set14\cite{set14}} & RDN\cite{rdn} & \lfi{34.01}/\lfi{0.9212} & \lfi{30.57}/\lfi{0.8468} & \lfi{28.81}/\lfi{0.7871} & - & -  \\
                          & RDN-MetaSR\cite{metasr} & 33.81/0.9194 & 30.43/0.8453 & 28.69/0.7858 & 26.43/0.6975 & 24.87/0.6370 \\
                          & RDN-LIIF\cite{liif} & 33.83/0.9196 &30.41/0.8456 &28.70/0.7857 &\lse{26.54}/0.7008 &\lse{25.05}/0.6446 \\
                        %   & RDN-\sysname (ours) & \lfi{34.15} & \lfi{30.59} & \lfi{28.87}  & \lfi{26.68} & \lfi{25.18} \\
                          & RDN-\sysname (ours) & 33.94/0.9201 & 30.47/0.8456 & 28.75/0.7870  & \lfi{26.58}/\lfi{0.7021} & \lfi{25.09}/\lfi{0.6462} \\
    \midrule
    \multirow{4}{*}{B100\cite{b100}} & RDN\cite{rdn} &\lfi{32.34}/\lfi{0.9017} & \lse{29.26}/0.8093 &27.72/0.7419 &- &- \\
                          & RDN-MetaSR\cite{metasr} & 32.29/0.9008 &\lse{29.24}/0.8091 &27.71/0.7414 &25.89/0.6522 &24.82/0.5977 \\
                          & RDN-LIIF\cite{liif} &32.28/0.9004 &29.25/0.8095 &27.73/0.7420 &\lse{25.97}/0.6555 & \lse{24.91}/0.6033 \\
                        %   & RDN-\sysname (ours) & \lfi{32.35} & \lfi{29.29} & \lfi{27.77}& \lfi{25.99} & \lfi{24.92} \\
                           
                          & RDN-\sysname (ours) & 32.31/0.9010 & \lfi{29.28}/\lfi{0.8098} & \lfi{27.76}/\lfi{0.7429} & \lfi{26.00}/\lfi{0.6563} & \lfi{24.93}/\lfi{0.6038} \\
    \midrule
    \multirow{4}{*}{Urban100\cite{urban100}} & RDN\cite{rdn} &32.89/0.9353 &28.80/0.8653 &26.61/0.8028 &- &- \\
                          & RDN-MetaSR\cite{metasr} &\lse{32.88}/0.9349 & \lse{28.85}/0.8660 &26.58/0.8020 &24.00/0.6943 &22.61/0.6205 \\
                          & RDN-LIIF\cite{liif} & 32.83/0.9344& \lse{28.80}/0.8656& \lse{26.67}/0.8037& \lse{24.20}/0.7033& \lse{22.79}/0.6348\\
                        %   & RDN-\sysname (ours) & \lfi{33.00} & \lfi{28.96} & \lfi{26.80} & \lfi{24.27} &  \lfi{22.83} \\
                          & RDN-\sysname (ours) & \lfi{32.97}/\lfi{0.9357} & \lfi{28.92}/\lfi{0.8675} & \lfi{26.76}/\lfi{0.8062} & \lfi{24.26}/\lfi{0.7061} &  \lfi{22.87}/\lfi{0.6374} \\
    \bottomrule
    \end{tabular}
    }
    \caption{Quantitative comparison on 4 benchmarks\cite{set5, set14, b100, urban100} with PSNR (dB) / SSIM on different scales. LIIF\cite{liif}, \sysname (ours) use a uniform model for all the scales. RDN\cite{rdn} uses different models for different scales and cannot be evaluated on out-of-distribution scales. The best results on different settings are bolded. }
    \label{tab:quantitative-benchmark}
    \vspace{-4mm}
\end{table*}

\section{Experiments}
In this section, we introduce the datasets and metrics in the experiments, the implementation details, and compared methods. Then we compare \sysname with some baseline alternatives both quantitatively and qualitatively. We further design experiments to discuss the generalization ability of \prefix and validate some of the design choices via ablation studies.  

\topic{Datasets and metrics}
We train all models using the DIV2K dataset~\cite{div2k} from NTIRE 2017 Challenge for all the experiments. This dataset consists of 1,000 HR images in 2K resolution and their corresponding LR images of different down-sampling scales ($\times 2 - \times 4$). We follow the original split setting, \ie 800 images for training, 100 images for validation. Consistent with previous works\cite{liif, metasr, rdn, edsr}, we also test our model on 4 popular benchmarks, \ie Set5\cite{set5}, Set14\cite{set14}, B100\cite{b100}, and Urban100\cite{urban100}, and report the results. We compare our method with baseline methods on several discrete scales ($\times 2, \times 3, \times 4$). To demonstrate that our method can be applied for arbitrary-scale SR, we also compare the SR results on a variety of large scales ($\times 6 - \times 30$) that do not appear in training time. We use the Peak Signal-to-Noise Ratio (PSNR) and SSIM\cite{ssim} as our evaluation metrics. Consistent with most SR works, the metrics are computed on RGB channels for DIV2K\cite{div2k} and Y channel (luminance)for other benchmarks\cite{set5, set14, b100, urban100}.

\topic{Training and inference details}
During training, we feed patches of size $48 \times 48$ into the network. To enable the network to perform arbitrary-scale SR on the input patches, we sample the SR scale $s$ from $\mathcal{U}(1,4)$. We then randomly crop a patch of size $48s \times 48s$ from the training HR image. The cropped HR patch is then down-sampled to $48 \times 48$ with bi-cubic interpolation as the LR input. For the implicit decoder training, we randomly select $48^2$ pixels from each HR patch. All the modules are trained from scratch with L1 loss\cite{edsr,rdn} for better convergence. We use Adam\cite{adam} optimizer with $\beta = (0.9, 0.999)$ to train the whole network for 1000 epochs. Each epoch contains 1000 iterations. In each iteration, we feed a batch of 16 sampled patches. The learning rate is initialized to ${10}^{-4}$ and decays to half every 200 epochs. For evaluation on scales within the distribution of training ($\times 2 \sim \times 4)$, we use the LR images provided in the DIV2K and different benchmark datasets. For out-of-distribution validation ($\times 6\sim\times 30$), we generate the LR inputs by resizing the HR images with bi-cubic interpolation.

\topic{Compared methods}
The methods involved in the experiments include: 1. models with CNN architectures targeted for a specific scale: EDSR-baseline\cite{edsr}, RDN\cite{rdn}, 2. the state-of-the-art arbitrary-scale SR methods at present: Meta-SR\cite{metasr}, LIIF\cite{liif}, 3. methods in group 2 that combines IPE: IPE-LIIF, IPE-MetaSR. For the latter two groups, both the feature extraction modules in EDSR-baseline and RDN could be adopted. In \cref{sec:result}, we evaluate all the models except IPE-MetaSR on benchmarks with various scales to verify the effectiveness of IPE on improving performance for arbitrary-scale SR methods. In \cref{sec:general}, we evaluate LIIF, MetaSR and their IPE-version after training on a wider range of SR scales to highlight the generality of IPE on multiple methods and larger training scales.

\subsection{Results and comparison}
\label{sec:result}

We compare \sysname with five baseline methods: bicubic interpolation, EDSR-baseline\cite{edsr}, RDN\cite{rdn}, MetaSR\cite{metasr} and LIIF\cite{liif}. Note that EDSR-baseline and RDN train different models for different scales, thus cannot be applied to out-of-distribution scales. The remaining parametric methods are designed for arbitrary-scale image SR and their feature extraction modules are exchangeable. The feature extraction module used by the three models is indicated by the prefix of the name. All the parametric methods are trained on the DIV2K\cite{div2k} training set. 

\topic{Quantitative comparison}
The quantitative results on the 5 benchmark datasets\cite{div2k, neighbor-embedding, set14, b100, urban100} are listed in \cref{tab:quantitative-div2k} and \cref{tab:quantitative-benchmark}. We observe that our method surpasses existing methods under almost all the test settings, especially on large datasets\cite{b100, urban100}. 
For SR on out-of-distribution scales, we achieve consistently better results than existing methods on all the benchmarks. 
In summary, the use of IPE enhances the existing SR model based on implicit representation, allowing them to go beyond previous scale-specific ones\cite{rdn} to support multiple scale factors.  We argue that the power of \sysname comes from the combination of IPE and implicit representation, which effectively fuse information from different scales and spatial locations.

\begin{figure*}[t]
\centering
\captionsetup[subfloat]{labelformat=empty,justification=centering}
 \setlength {\tabcolsep} {2pt}
\begin{tabular}{cccc}
\multirow{-7}{*}{\subfloat[barbara from Set14\cite{set14} ($\times 2$)]{\includegraphics[width=0.45\linewidth]{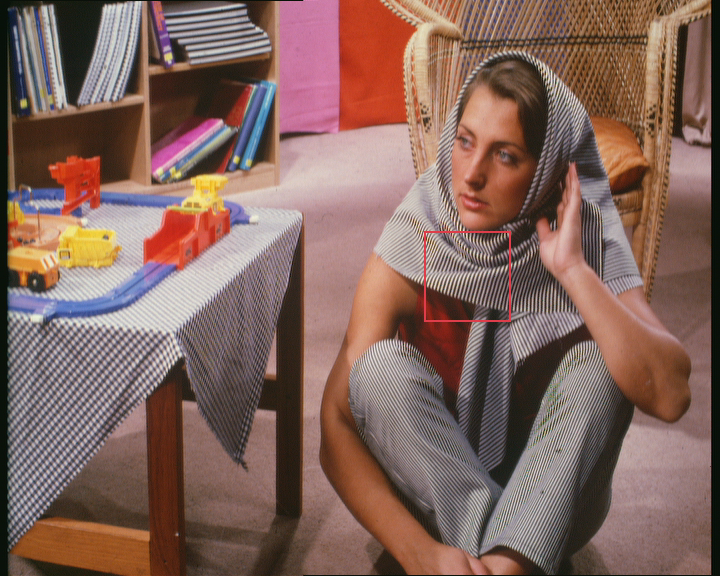}}} & 

\subfloat[HR\\(PSNR/SSIM)]{\includegraphics[width=0.15\linewidth]{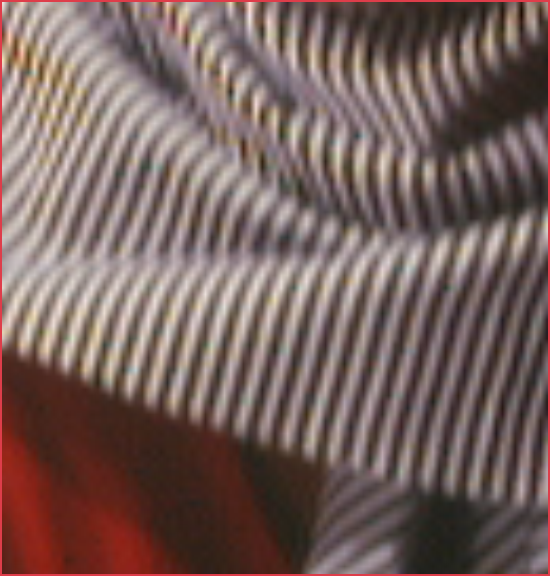}} & 
\subfloat[EDSR-LIIF\\(29.22dB/0.8768)]{\includegraphics[width=0.15\linewidth]{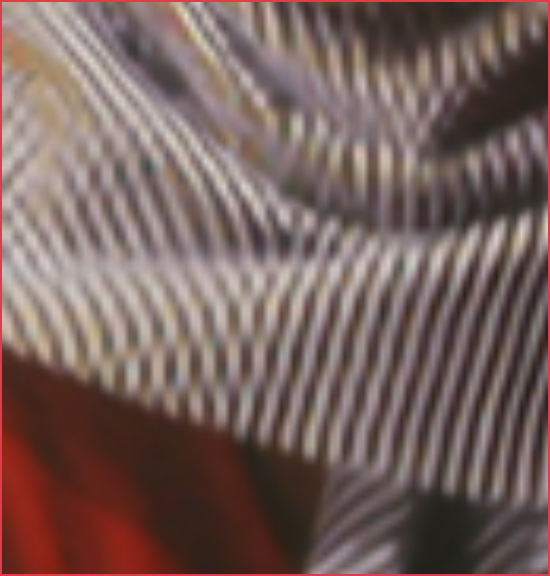}} & 
\subfloat[EDSR-IPE-LIIF\\(29.30dB/0.8770)]{\includegraphics[width=0.15\linewidth]{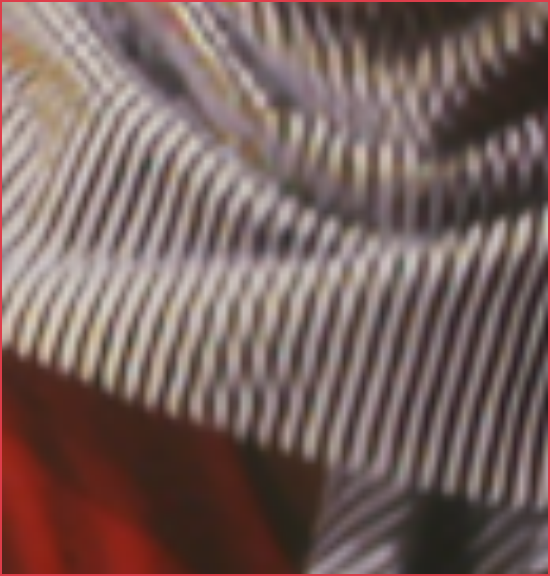}} \\
& 
\subfloat[Bicubic\\(27.94dB/0.8210)]{\includegraphics[width=0.15\linewidth]{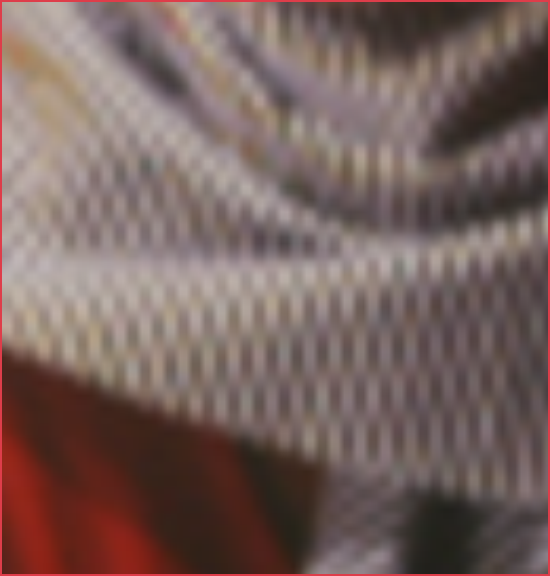}} & 
\subfloat[RDN-LIIF\\(30.07dB/0.8905)]{\includegraphics[width=0.15\linewidth]{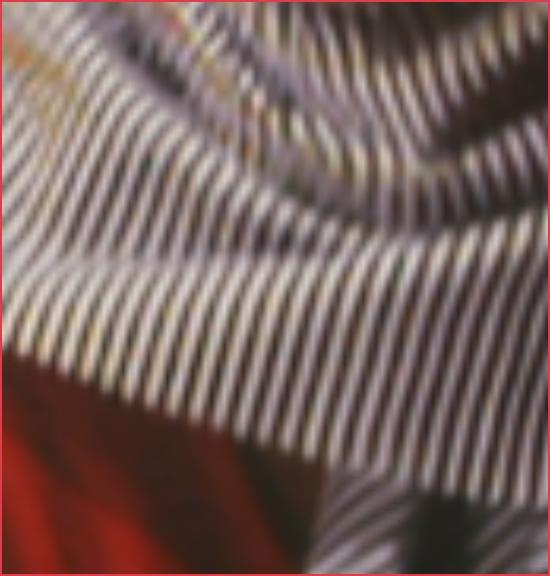}} & 
\subfloat[RDN-IPE-LIIF\\(30.16dB/0.8926)]{\includegraphics[width=0.15\linewidth]{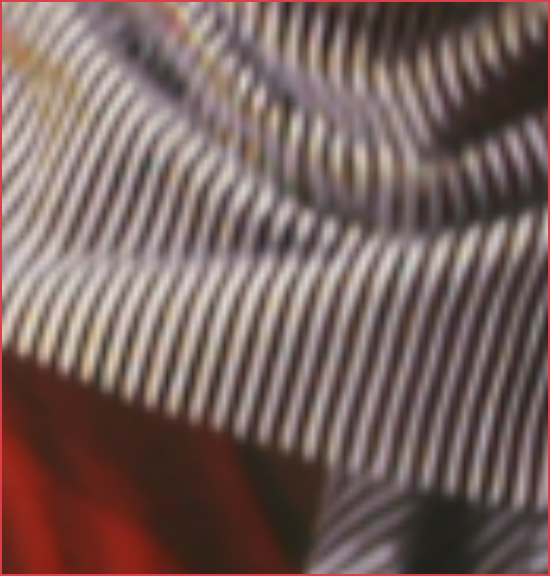}} \\

\multirow{-6.5}{*}{\subfloat[img096 from Urban100\cite{urban100} ($\times 4$)]{\includegraphics[width=0.45\linewidth]{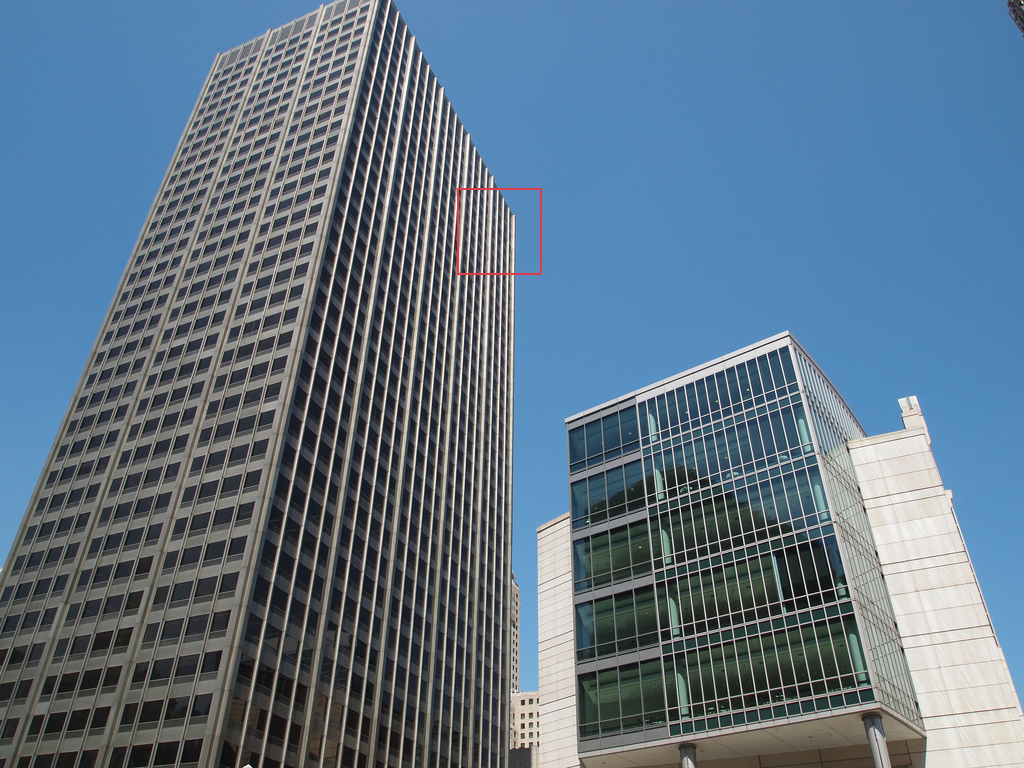}}} & 

% \subfloat[HR\\(PSNR/SSIM)]{\includegraphics[width=0.15\linewidth]{figure/result/cropped/HR_0095_pred_magnifier_0.png}} & 
% \subfloat[EDSR-LIIF\\(24.95dB)]{\includegraphics[width=0.15\linewidth]{figure/result/cropped/urban100-4-edsr-liif_0095_pred_magnifier_0.png}} & 
% \subfloat[EDSR-IPE-LIIF\\(25.36dB)]{\includegraphics[width=0.15\linewidth]{figure/result/cropped/urban100-4-edsr-ipe_0095_pred_magnifier_0.png}} \\
% & 
% \subfloat[Bicubic\\(21.38dB)]{\includegraphics[width=0.15\linewidth]{figure/result/cropped/X4_0095_pred_magnifier_0.png}} & 
% \subfloat[RDN-LIIF\\(26.10dB)]{\includegraphics[width=0.15\linewidth]{figure/result/cropped/urban100-4-rdn-liif_0095_pred_magnifier_0.png}} & 
% \subfloat[RDN-IPE-LIIF\\(26.57dB)]{\includegraphics[width=0.15\linewidth]{figure/result/cropped/urban100-4-rdn-ipe_0095_pred_magnifier_0.png}} \\

\subfloat[HR\\(PSNR/SSIM)]{\includegraphics[width=0.15\linewidth]{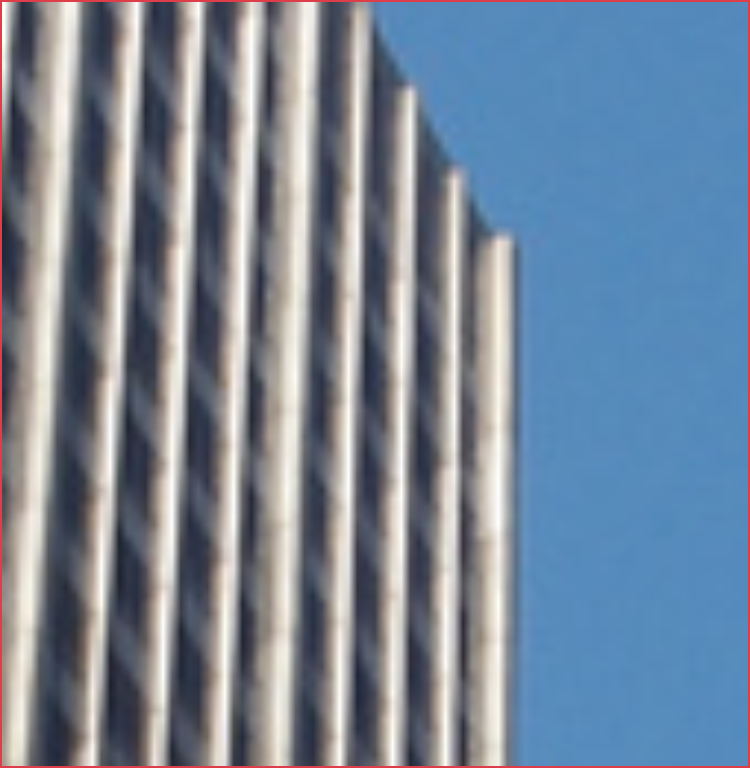}} & 
\subfloat[EDSR-LIIF\\(24.95dB/0.8513)]{\includegraphics[width=0.15\linewidth]{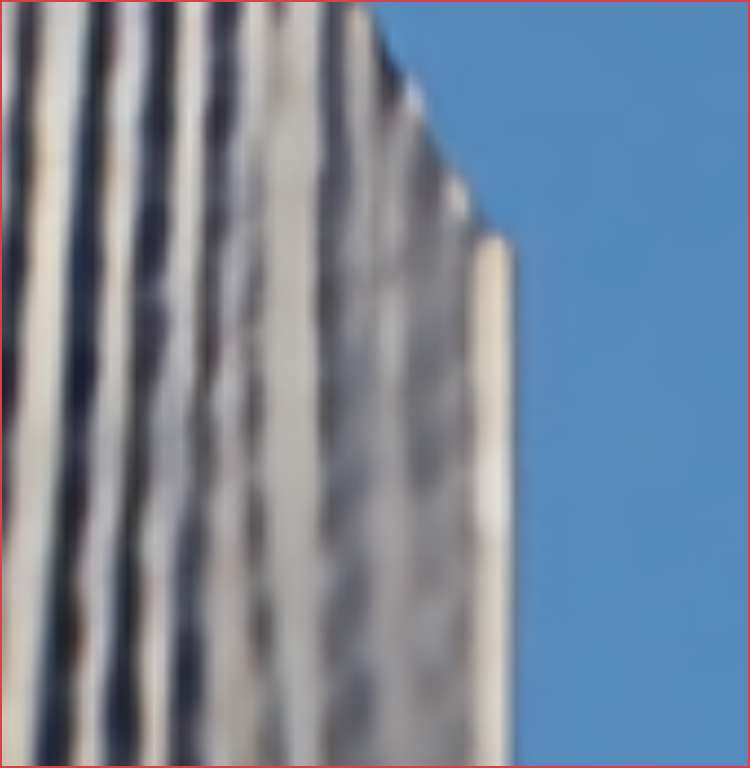}} & 
\subfloat[EDSR-IPE-LIIF\\(25.36dB/0.8603)]{\includegraphics[width=0.15\linewidth]{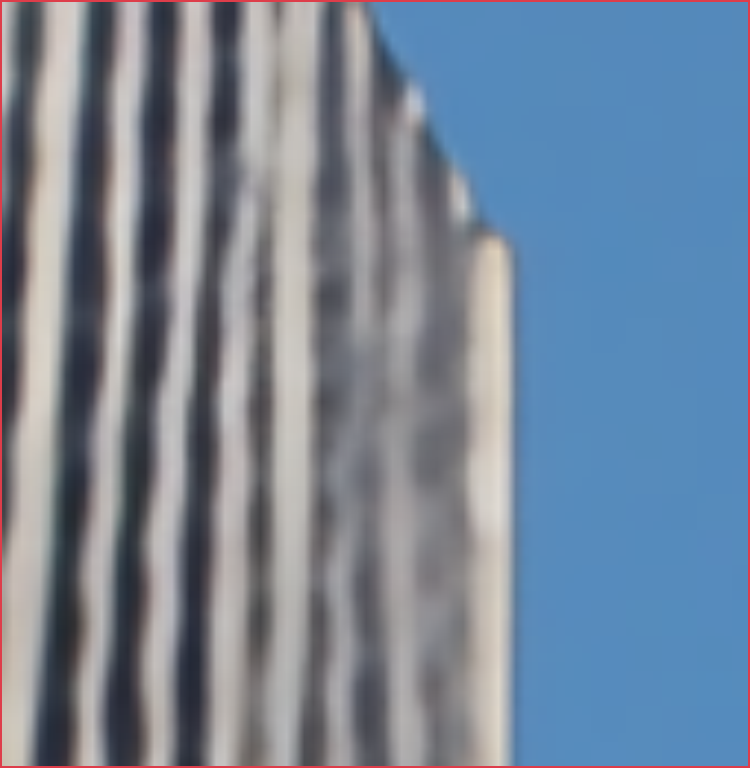}} \\
& 
\subfloat[Bicubic\\(21.38dB/0.6705)]{\includegraphics[width=0.15\linewidth]{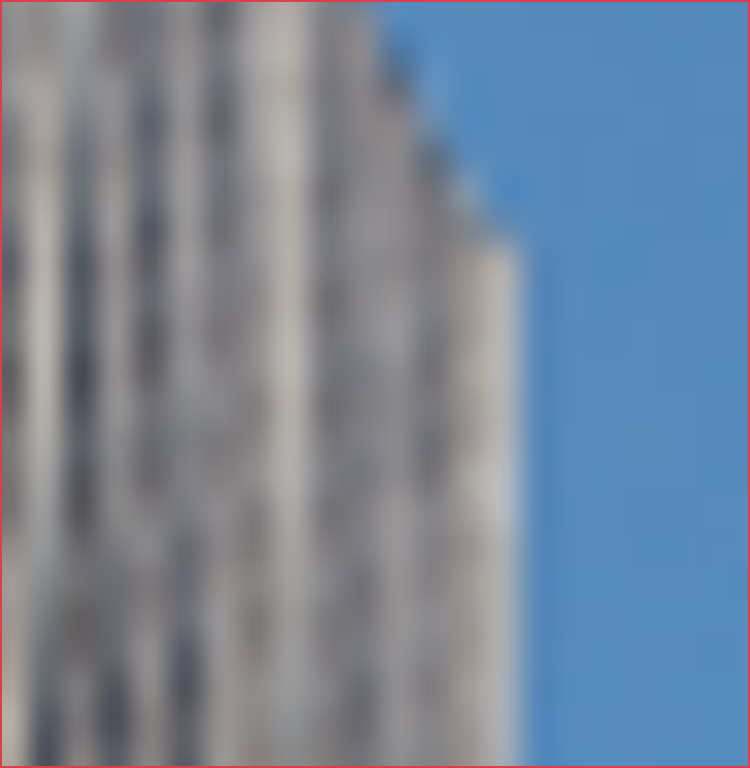}} & 
\subfloat[RDN-LIIF\\(26.10dB/0.8808)]{\includegraphics[width=0.15\linewidth]{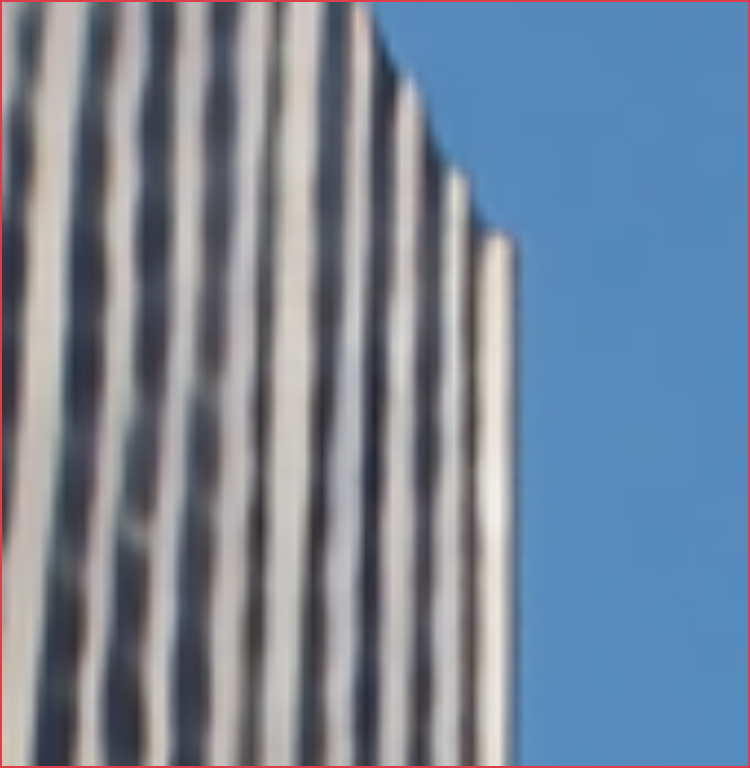}} & 
\subfloat[RDN-IPE-LIIF\\(26.57dB/0.8845)]{\includegraphics[width=0.15\linewidth]{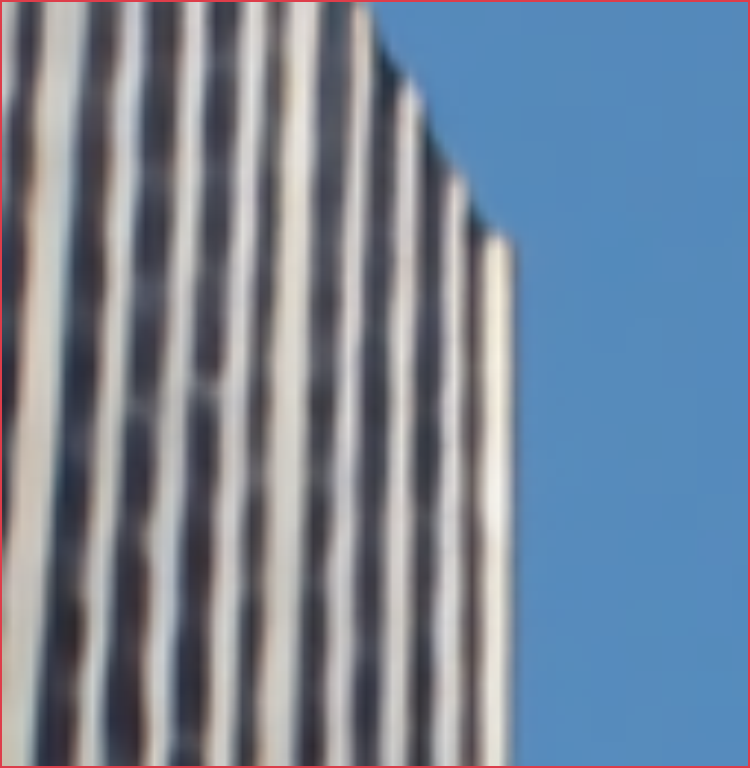}} \\

\multirow{-7}{*}{\subfloat[img042 from Urban100\cite{urban100} ($\times 4$)]{\includegraphics[width=0.45\linewidth]{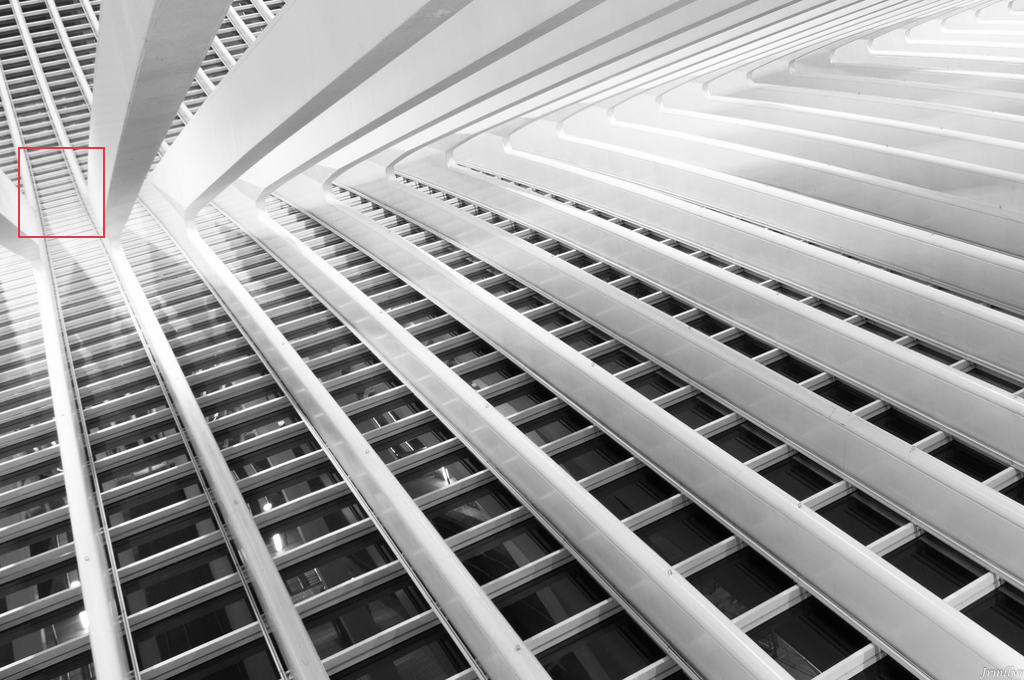}}} & 

\subfloat[HR\\(PSNR/SSIM)]{\includegraphics[width=0.15\linewidth]{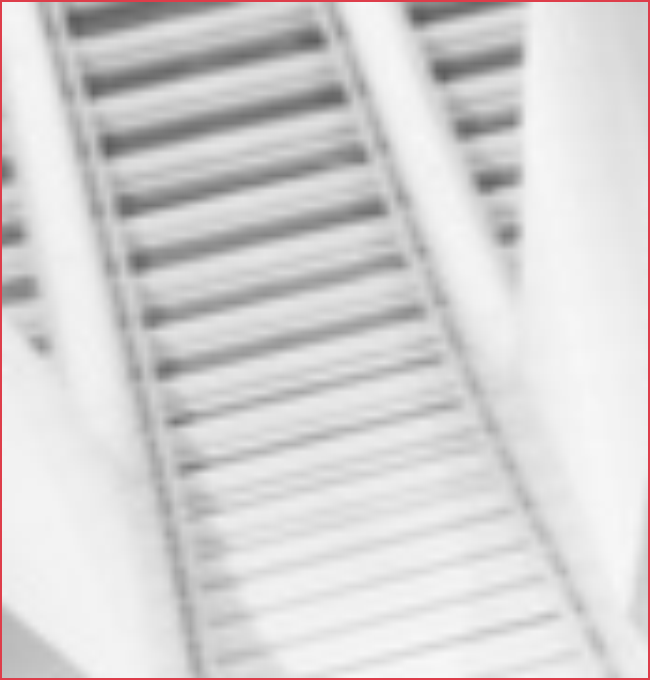}} & 
\subfloat[EDSR-LIIF\\(29.13dB/0.8842)]{\includegraphics[width=0.15\linewidth]{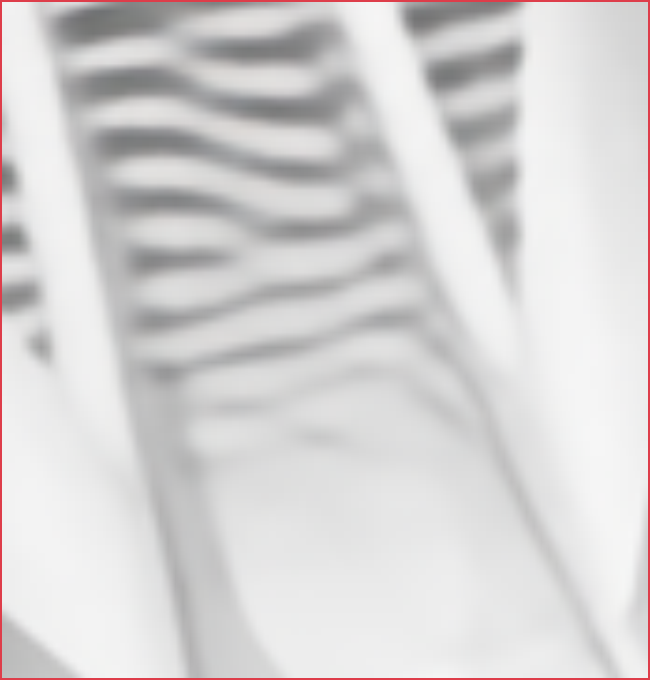}} & 
\subfloat[EDSR-IPE-LIIF\\(29.33dB/0.8880)]{\includegraphics[width=0.15\linewidth]{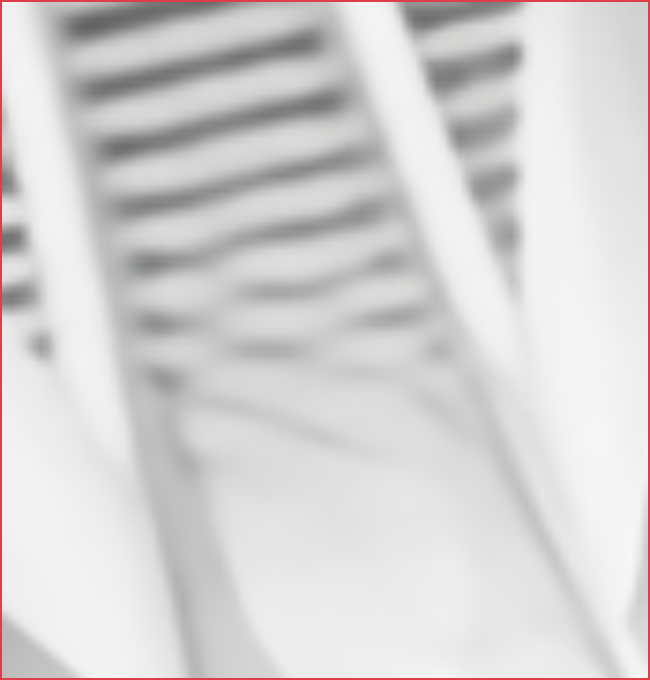}} \\
& 
\subfloat[Bicubic\\(21.90dB/0.6798)]{\includegraphics[width=0.15\linewidth]{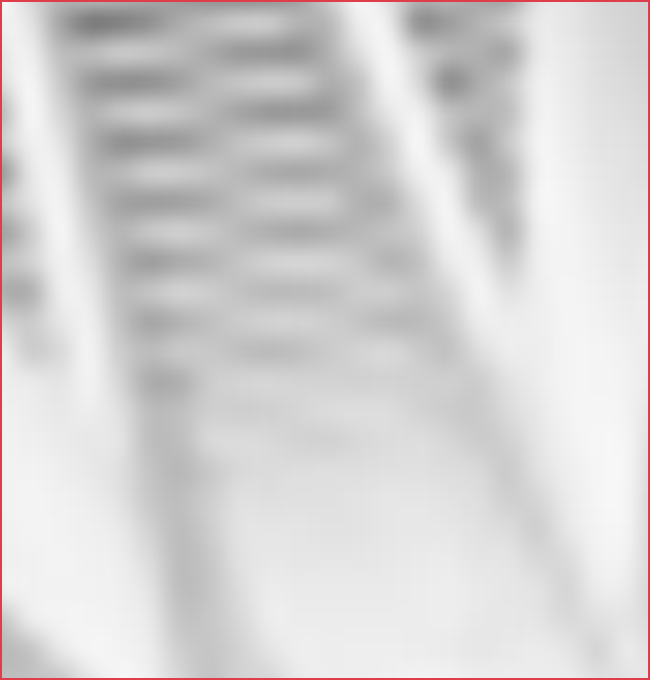}} & 
\subfloat[RDN-LIIF\\(29.41dB/0.8906)]{\includegraphics[width=0.15\linewidth]{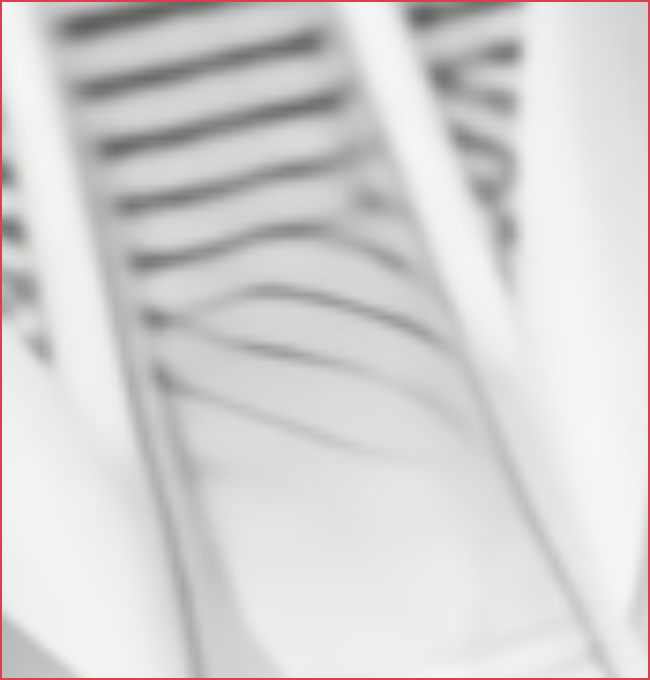}} & 
\subfloat[RDN-IPE-LIIF\\(29.67dB/0.8964)]{\includegraphics[width=0.15\linewidth]{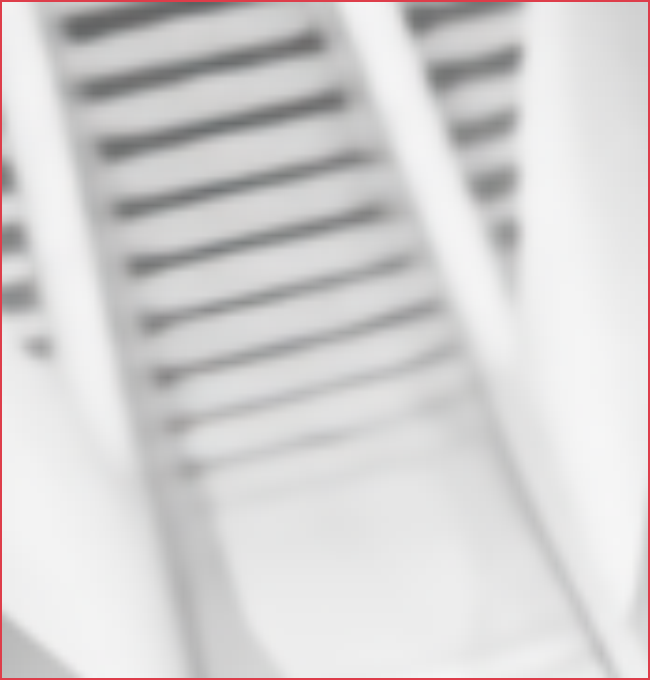}} \\

\end{tabular}
\caption{Visual comparison of bicubic, LIIF, and \sysname on different SR scales.}
\label{fig:qualitative-comparison}
\end{figure*}

\topic{Qualitative comparison}
In \cref{fig:teaser_example} and \cref{fig:qualitative-comparison}, we further show visual image comparisons of the previous work and our method on different SR scales. For areas containing scale variant stripes, the previous method\cite{liif} will produce checkboard artifacts and could not faithfully recover the HR contents, while \sysname can distinguish texture details by encoding the query pixel spatial information. In these cases, IPE is able to better identify the position and size of complex textures so that they can be clearly distinguished.

\subsection{General enhancement for multi-scale implicit learning}
\label{sec:general}
Both Meta-SR\cite{liif} and LIIF\cite{liif} take the advantage of the implicit representation. The former predicts an up-scaling filter for each output pixel and the latter predicts the color directly with a coordinate input via implicit decoding. The IPE strategy can also be easily applied to improve Meta-SR. That is, for a given pixel $(i, j)$ on the SR image output up-scaled by $s$, we redefine the original input to  Meta-Upscale Module\cite{metasr}: 
\begin{equation}
\mathbf{v}_{ij} = (\frac i s -\lfloor \frac i s \rfloor, \frac j s - \lfloor \frac j s \rfloor, \frac 1 s)
\end{equation}
 as
 \begin{equation}
 \mathbf{v}_{ij} = \hat{\gamma}((\frac i s -\lfloor \frac i s \rfloor, \frac j s - \lfloor \frac j s \rfloor), (\frac 1 s, \frac 1 s)) 
 \end{equation} and form IPE-MetaSR. 
To verify the effectiveness of IPE on a relatively large range of scale variations, \eg $\mathcal{U}(1,8), \mathcal{U}(1,16)$, we train several candidate models and evaluate them on DIV2K\cite{div2k} over the same scale distribution. From \cref{tab:general-enhancement}, we can observe that with IPE, both models obtain consistent improvements on different SR scales. By comparing the results on large scales presented in \cref{tab:general-enhancement} and \cref{tab:quantitative-div2k}, it can be seen that LIIF fails to exceed the performance of our method without these training samples even if the data on $\times 6 - \times 12$ is provided. This experiment illustrates the generality of IPE on multiple implicit models and a wider range of SR scales.
 
 \begin{table}[tbp]
    \centering
     \resizebox{\linewidth}{!}{
    \begin{tabular}{c|c|ccccc}
    \toprule
    
     Scale  & Method & $\times 2$ & $\times 3$& $\times 4$ & $\times 6$& $\times 12$\\
     \midrule
     \multirow{6}{*}{$\mathcal{U}(1,8)$} 
     & LIIF\cite{liif} & \slse{34.60} & \slse{30.94} &	\slse{29.01} &	26.79 & -\\
     & LIIF w/ PE & 34.47 &	30.87 &	\slse{29.01}	& \lfi{26.81} & -\\
     & ipe-LIIF(ours) & \lfi{34.65}	& \lfi{30.98} &	\lfi{29.04}	& \lfi{26.81} & -\\
     \cmidrule(lr){2-7}
     & MetaSR\cite{metasr} & 34.57 &	30.91	& 28.97 &	26.72 & -\\
     & MetaSR w/ PE & \slse{34.60} &	\lfi{30.93}	& \lfi{28.99} &	\lfi{26.74} & -\\
     & ipe-MetaSR(ours) & \lfi{34.61} & \lfi{30.93}	& \lfi{28.99} & \lfi{26.74} & -\\
     \midrule
     \midrule
     \multirow{6}{*}{$\mathcal{U}(1,16)$}  
     & LIIF\cite{liif} & \slse{34.50}	& \slse{30.89}	& \slse{28.97}	&26.76	& 23.75 \\
     & LIIF w/ PE & 34.36 &	30.79 &	28.95	& \slse{26.78} &	\lfi{23.79} \\ 
     & IPE-LIIF(ours) & \lfi{34.54}	& \lfi{30.92} &	\lfi{29.00}	& \lfi{26.79} &	\slse{23.76} \\
    %  \midrule
     \cmidrule(lr){2-7}
    &  MetaSR\cite{metasr} & 34.48 & \lfi{30.86} & \lfi{28.94} & 26.72	& 23.70 \\
    & MetaSR w/ PE & 34.50 & 	30.85	& 28.93 &	26.72	& 23.70 \\
     & IPE-MetaSR(ours) & \lfi{34.51} & \lfi{30.86} & 	\lfi{28.94} & 	\lfi{26.74} &	\lfi{23.71} \\
     
    \bottomrule
    \end{tabular}}
    \caption{Validation results of LIIF\cite{liif} and Meta-SR\cite{metasr} on DIV2K\cite{div2k} combined with EDSR-baseline and different coordinate encodings, including none, plain positional encoding, and integrated positional encoding. The best results are bolded.}
    \label{tab:general-enhancement}
    \vspace{-4mm}
\end{table}

\subsection{Ablation studies}
\label{sec:ablation}
We discuss the effectiveness of some design choices of \sysname in this section, including: 
1. the bandwidth $L$ of integrated positional encoding,
2. whether to use the integrated version of PE,
3. whether to use the original cell decoding, 
4. whether to use skip connections in the implicit decoder.  
EDSR-baseline is used as the feature extraction module for all the ablation studies. The quantitative results are listed in \cref{tab:ablation} and green indicates a decrease in results.

\topic{Bandwidth $L$ of integrated positional encoding} In terms of bandwidth $L$, we choose $L = 10$ in \sysname and set $L = 4, 16$ for competitors. Comparing the two variants $(L = 4, 16)$ with the original model, we observe that the effect of the bandwidth is minor. Taking the results and computational overhead into account, we choose a mild value $L = 10$. The intuitive explanation of how a large $L$ does not improve the SR quality is that the additional terms in IPE (\cref{eq:ipe}) have large denominators and remain almost constant near 0 as the query pixel region changes. That is, a larger $L$ does not necessarily provide valid spatial information, which is also highlighted in mip-NeRF\cite{mip-nerf}. 

\begin{table}[htbp]
    \centering
    \resizebox{\linewidth}{!}{
    \begin{tabular}{l|ccc|ccc}
    \toprule
    \multirow{2}{*}{Method} & \multicolumn{3}{c|}{In-distribution} & \multicolumn{3}{c}{Out-of-distribution} \\
       & $\times 2$ & $\times 3$& $\times 4$ & $\times 6$&
       %$\times 12$& 
       $\times 18$
       %$\times 24$ 
       & $\times 30$\\
    \midrule
   \sysname(ours) & 34.72 & {31.01}	& {29.04}	& {26.79} 
   %& {23.75} 
   & {22.21} 
   %& {21.22}	
   & {20.51} \\
    \sysname (\textit{L} = 16) &  34.72 &	31.01 &	29.04 &	26.79 
    %& 23.76 
    & 22.21	
    %& 21.22
    & 20.52 \\
    \sysname (\textit{L} = 4) & 34.71 &	31.00 &	29.03 &	26.78	
    %& 23.75 
    & 22.21	 
    % & 21.22 
    &	20.52 \\
    LIIF w/ PE (\textit{L} = 10) & \down{34.59}	& \down{30.92} & \down{29.00} &	\down{26.76}	
    % & 23.74 
    &	22.21 
    % & 21.22	
    & 20.52 \\
    \sysname + cell & 34.72 & 31.01	& 29.04 &	26.79	
    %& 23.74 
    &	\down{22.18}	
    %& \down{21.18} 
    & \down{20.49} \\
    \sysname - skip & \down{34.67}	& \down{30.97} &	\down{29.01} &	\down{26.76} 
    %&	\down{23.72} 
    &	\down{22.19} 
    %&	\down{21.20} 
    &	20.50\\
    
    \bottomrule
    \end{tabular}
    }
    \caption{Ablation of design choices of \sysname on DIV2K\cite{div2k} validation set. EDSR-baseline is used as the feature extraction module for all the settings. The results with degraded performance are marked in green.}
    \label{tab:ablation}
    \vspace{-4mm}
\end{table}
\topic{Integrated positional encoding} 
An obvious possible improvement on LIIF\cite{liif} is to add plain PE to the coordinate input. To verify the effectiveness of IPE, we perform plain PE on the input and form LIIF with PE. For a fair comparison, we also set the bandwidth $L$ to 10. It can be seen that with PE, the network focuses more on high-frequency information and performs poorly on relatively small scale factors, sometimes even worse than the original LIIF. 

\topic{Other design choices}
In \cref{tab:ablation}, we show the ablation results on some other design choices, including the use of cell decoding and skip connections in the implicit decoder. We concatenate the cell size with the IPE input and form \sysname + cell. It can be seen that cell decoding makes the results on large scaling factors worse, which is because it is not able to generalize to arbitrary scale factor. For the out-of-distribution scales, the cell code is relatively small compared with those seen in training time. We also remove skip connections and form \sysname - skip. The results verify that SR on almost all scales benefits from the skip connections, as it provides a more flexible network depth.

From \cref{tab:ablation}, we find that IPE performs good aggregation of the spatial information in the query region. It combines the advantages of PE and cell decoding and provides a universal solution to implicit learning tasks with multi-scale samples.

\section{Limitations}
Although we have improved the existing arbitrary-scale image SR methods with IPE, regression-based methods still cannot fill in the missing details when the scale factor is large. In contrast, generative methods\cite{sr3, srgan, esrgan} could induct reasonable semantic fill-in and produce visually clear results with a poor PSNR but high perceptual similarity\cite{lpips}. In addition, in \sysname, the wider layers in skip-MLP will increase the training time by 10\% to 20\%.

\section{Conclusion}
 In this paper, we propose to use IPE for arbitrary-scale image SR, and apply it to the state-of-the-art method LIIF. With the enhancement provided by IPE, \sysname generates consistently high-fidelity SR results on various scales, which is verified by quantitative and qualitative experiments.  Further experiments on the generalization ability suggest that IPE can be seamlessly integrated to multiple architectures to increase the capacity of implicit learning on multi-scale samples.

{\small
\bibliographystyle{ieee_fullname}
\bibliography{egbib}
}

\end{document}